\documentclass[twocolumn, pra, superscriptaddress, amsmath, amssymb, aps, showpacs]{revtex4}

\usepackage{verbatim}
\usepackage{amsmath,amssymb,mathrsfs}
\usepackage{cases}
\usepackage{graphicx}
\usepackage{dcolumn}
\usepackage{bm}
\usepackage{color}
\pagecolor{white}
\usepackage{amsfonts,latexsym}
\usepackage{enumerate}
\usepackage{rotating}
\usepackage{epstopdf}
\usepackage{braket}
\usepackage{hyperref}
\hypersetup{
	colorlinks=true,
	linkcolor=blue,
	filecolor=blue,
	urlcolor=blue,
	citecolor=magenta
}

\newcommand{\beq}{\begin{equation}}
	\newcommand{\eeq}{\end{equation}}
\newcommand{\beqa}{\begin{eqnarray}}
	\newcommand{\eeqa}{\end{eqnarray}}
\newcommand{\Beqa}{\begin{eqnarray*}}
	\newcommand{\Eeqa}{\end{eqnarray*}}

\def\bal#1\eal{\begin{align}#1\end{align}}
\def\Bal#1\Eal{\begin{align*}#1\end{align*}}

\begin{document}

\title{Dynamical transition of the generalized Jaynes–Cummings model: multi-particles and inter-particle interaction effects}

\author{Wen Liang}
\affiliation{Guangdong Provincial Key Laboratory of Quantum Metrology and Sensing, and School of Physics and Astronomy, Sun Yat-Sen University (Zhuhai Campus), Zhuhai 519082, China}

\author{Zhenhua Yu}
\email[]{huazhenyu2000@gmail.com}
\affiliation{Guangdong Provincial Key Laboratory of Quantum Metrology and Sensing, and School of Physics and Astronomy, Sun Yat-Sen University (Zhuhai Campus), Zhuhai 519082, China}
\affiliation{State Key Laboratory of Optoelectronic Materials and Technologies, Sun Yat-Sen University (Guangzhou Campus), Guangzhou 510275, China}

\date{\today}

\begin{abstract} 
How environments affect dynamics of quantum systems remains a central question in understanding transitions between quantum and classical phenomena and optimizing quantum technologies. A paradigm model to address the above question is the generalized Jaynes–Cummings model, in which a two-level particle is coupled to its environment modeled by a continuum of boson modes. Previous analytic solutions show that, starting from the initial state that the particle is in its excited state and the boson modes in their vacuum state, the time evolution of the probability that the particle occupies the excited state exhibits a dynamical transition as the system-environment coupling varies; when the coupling is weak, the probability decays to zero monotonically, while a finite weight of the particle is localized in the excited state when the coupling is sufficiently strong. Here, we study the dynamical transition for the case that $N$ particles are initially excited with the boson modes in their vacuum state. The boson modes are assumed to follow a spectral function $J(\omega)\sim\omega^s$ up to a cut-off frequency $\omega_c$ with $s>0$. In particular, we access the effects of an all to all Ising type interaction we introduce between the particles. Our calculation is carried out by the non-perturbative numerical renormalization group method. We find that the value of the critical coupling for the transition exhibits a maximum at a finite $N$ and decreases with sufficiently large $N$, and is suppressed (enlarged) by a ferromagnetic (anti-ferromagnetic) Ising interaction. Our results enrich understanding on environmental effects on interacting quantum systems.
\end{abstract}

\maketitle


\section{Introduction}
Quantum dynamics lies at the center of the research of quantum science; compared with classical dynamics, its peculiarity offers novel prospective applications in quantum technologies \cite{Galindo2002,Degen2017,Treutlein2018,Amico2022}. 
However, since experiments inevitably subject quantum systems to environments, it is crucial to investigate environmental effects on dynamical processes such as decoherence, dissipation, and entanglement \cite{dobrovitski2003quantum,yao2007restoring,hanson2008coherent,orth2010dynamics,strathearn2018efficient,2021Quantum}, which often constrain the robustness of intended quantum operations.


A prototype model addressing the problem of decoherence due to a dissipative environment is the spin-boson model \cite{leggett1987dynamics,weiss2012quantum}, whose Hamiltonian is usually given by $H_{\rm S-B}=\Delta\sigma_x/2+\sigma_z\sum_\nu\lambda_\nu (a_\nu+a^\dagger_\nu)+\sum_\nu\omega_\nu a_\nu^\dagger a_\nu$. This model derives from the familiar problem of a single particle tunneling between the two minima in a double well potential \cite{weiss1987incoherent,jelic2012double}. The two eigenstates of $\sigma_z$, $\ket\uparrow$ and $\ket\downarrow$, correspond to the left and right minimum of the potential respectively. The off-diagonal term $\sim\Delta\sigma_x$ brings about the oscillation between the two eigenstates, equivalent to the particle tunneling between the two minima. The dissipative environment is modeled by a continuum of boson modes whose creation (annihilation) operators are $a_\nu^\dagger$ ($a_\nu$). 
The coupling between the spin and the bosons indicates that the environment constantly ``monitors" the state (position) of the spin (particle). Thus it is anticipated if such a monitoring is strong enough, the environment collapses the spin state into either $\ket\uparrow$ or $\ket\downarrow$, and wipes out the quantum coherence. 

It has been shown in Ref.~\cite{leggett1987dynamics} that the environmental effects on the spin dynamics are encapsulated in the spectral function $J(\omega)$ of the bosons [see Eq.~(\ref{j0})]; at zero temperature, for $\Delta\to0$, the spin dynamics is localized in the eigenstates of $\sigma_z$ for sub-Ohmic $J(\omega)$, and undergoes a damped oscillation for super-Ohmic $J(\omega)$, and transits from a damped oscillation to an incoherent relaxation and to localization for Ohmic $J(\omega)$ with increasing magnitude. Furthermore, for finite $\Delta$, numerical renormalization group calculations 
observed coherent dynamics even for sub-Ohmic $J(\omega)$ \cite{bulla2003numerical,anders2007equilibrium}, and 
the numerically exact time-evolving matrix product operator method identified a new phase characterized by an aperiodic behavior for strong coupling \cite{otterpohl2022hidden}. 
The spin-boson model 
not only sheds light on the fundamental question of how quantum phenomena transit to classical behavior \cite{breuer2002theory,ingold2002path}, but also provides a base to study topics ranging from electron transport to quantum information \cite{zurek2003decoherence,kapral2015quantum,de2021colloquium,cai2023quantum,hangleiter2023computational,schuster2023operator}.

Beyond the single-spin case, further works continued to investigate multi-spin cases, mostly with inter-spin interactions \cite{dube1998dynamics,governale2001decoherence,thorwart2001decoherence,garst2004quantum,nagele2008dynamics,nagele2010dynamics,orth2010dynamics,mccutcheon2010separation,bonart2013dissipative,vorrath2004dynamics,vorrath2005dynamics,anders2008dynamics,winter2014quantum,
werner2005quantum,cugliandolo2005static,schehr2006strong,schehr2008finite,hoyos2008theory,hoyos2012dissipation,de2020quantum,2021Quantum}. 
A common bath that a pair of spin couple to is found to mediate an indirect interaction between the spins, giving rise to both mutual coherence effects and dissipation \cite{dube1998dynamics}. When the pair of spins are subject to a CNOT gate, the gate operation is sensitive to the coupling strength with the bath \cite{thorwart2001decoherence}. Exact results of spin dynamics is available in the white noise limit \cite{nagele2008dynamics,nagele2010dynamics}. Particular nonequilibrium initial preparations are capable of making the system steady states hightly entangled \cite{orth2010dynamics}. Under certain circumstances, a finite number of spins coupled to a common bath can be mapped to a giant spin of magnitude $J$ coupled to the bath \cite{vorrath2004dynamics,vorrath2005dynamics,anders2008dynamics}; at zero temperature, in the absence of direct spin interactions, the quantum phase transition is shown to be in the same universality class as the single spin-boson model \cite{winter2014quantum}.
When the number of spins is elevated toward the thermodynamic limit, Monte Carlo simulations indicate that the quantum phase transition in a one-dimensional Ising model in a transverse magnetic field coupled to a dissipative heat bath is so different from the single spin-boson model \cite{werner2005quantum} that a new quantum criticality is formed regardless of dissipation strength. If the ferromagnetic couplings of the spin chain are set random, the original sharp transition is smeared \cite{cugliandolo2005static,schehr2006strong,schehr2008finite,hoyos2008theory,hoyos2012dissipation}.

It is known that different specific coupling forms between systems and environments can lead to drastically different behaviors in system dynamics \cite{breuer2002theory,zhang2023dynamical}. Other than the spin-boson model, another widely studied model is the Jaynes–Cummings model generalized to a continuum boson modes. The two models are related in the sense that the Hamiltonian of the latter is $\sigma_y H_{\rm S-B}\sigma_y$ with only the system-environment coupling terms $\sim\sigma^+ a_\nu$ and $\sigma^- a_\nu^\dagger$ retained. It can be shown analytically that if initially the particle is in its excited state and the bosons are in their vacuum state, the time evolution of the probability that the particle continues occupying the excited state exhibits a dynamical transition \cite{Pfeifer1982,cohen1998atom}: when the system-environment coupling is weak, the probability decays to zero monotonically; when the coupling is strong, a finite weight of the particle is localized in the excited state. The generalized Jaynes–Cummings model has been employed to explain stability of certain molecule states \cite{Pfeifer1982,Bahrami2011,Taher2013}, and to explore non-Markovian behavior for a particle embedded in a boson bath \cite{John1994,Florescu2001,Burgess2022}. However, our knowledge of the model system when applied to multiple particles, especially with interaction included, is rather limited.

In this work, we study the dynamical transition of the generalized Jaynes–Cummings model applied to multiple particles and a continuum boson bath. The boson bath is assumed to follow a spectral function $J(\omega)\sim\omega^s$ with $s>0$. We access the effects of an all to all Ising type inter-particle interaction on the transition [see Eq.~(\ref{h})]. Recently, this interaction is of particular interest partly due to its application in realizing the CNOT gate in quantum computation \cite{mermin2007quantum}.
Our results are calculated by the non-perturbative numerical renormalization group method \cite{anders2006spin}. Our numerical calculation agrees well with the benchmark provided by the analytic result for a single particle \cite{Pfeifer1982,cohen1998atom} (see Appendix C). The purpose of our study is to investigate the consequences of multi-particles and inter-particle interactions, and to explore their implications to quantum simulations and quantum computations.
 In the case of the number of particles $N\ge2$, we find that the critical value of the system-environment coupling required for the transition exhibits a maximum at a finite $N$, and decreases with sufficiently large $N$. Moreover, we find that the ferromagnetic (anti-ferromagnetic) Ising interaction suppresses (enlarges) the critical coupling value. Our main results are summarized in Figs.~\ref{2pe}-\ref{ac}.
Our results enrich the understanding of environmental effects on systems with intra-interactions.

\section{model}
We consider $N$ identical two-level particles coupled to a continuum of boson modes. 
The energy difference $E_a$ between the two internal levels of each particle $|e\rangle$ and $|g\rangle$ is close to the boson mode frequencies $\{\omega_\nu\}$. We take $\hbar=1$ throughout. 
We denote $\omega_l$ and $\omega_l+\omega_c$ as the lower bound and the upper bound of $\{\omega_\nu\}$.
It is convenient for us to choose $\omega_l$ as the zero point for energies.
Meanwhile, we consider an all to all Ising type interaction between the particles. This interaction can be used to realize the CNOT gate in quantum computation \cite{mermin2007quantum}.
The Hamiltonian of the combined system is given by 
\begin{align}
\tilde H=&\sum_{j=1}^N\frac{\Delta}2\sigma_j^z+g\sum_{j<k}\sigma_j^z \sigma_k^z+\sum_{\nu}\tilde\omega_\nu a_\nu^\dagger a_\nu+H_{p-b},\label{h}\\
H_{p-b}=&\sum_{j=1}^N\sum_{\nu}\lambda_\nu(\sigma_j^- a_\nu^\dagger+\sigma_j^+ a_\nu),
\end{align}
where $\Delta\equiv E_a-\omega_l$ and $\tilde\omega_\nu\equiv\omega_\nu-\omega_l$, and $\sigma_{j}^z=(|e_j\rangle\langle e_j|-|g_j\rangle\langle g_j|)/2$, $\sigma_j^+=|e_j\rangle\langle g_j|$ and $\sigma_j^-=|g_j\rangle\langle e_j|$, and 
$a_\nu^\dagger(a_\nu)$ are creation (annihilation) operators of the boson modes with frequency $\omega_\nu$,
$\lambda_\nu$ are the couplings between the particles and the boson modes, and $g$ is the inter-particle interaction coupling. The ferromagnetic coupling corresponds to $g<0$ and the anti-ferromagnetic one to $g>0$. The Hamiltonian, Eq.~(\ref{h}), is a generalization of the Jaynes–Cummings model  to multiple particles and boson modes, plus the inter-particle interaction introduced. We take the boson modes as a continuum bath.

As in the case for the renowned spin-boson model, the effects of the boson bath on the particle dynamics take place via the spectral distribution \cite{leggett1987dynamics,weiss2012quantum}
\begin{align}
J(\omega)=\pi\sum_{\nu}\lambda_\nu^2\delta(\omega-\tilde\omega_\nu);\label{j0}
\end{align}
this can be clearly seen if one integrates out the boson bath in the path integral formalism. 
We consider the continuum limit such that $J(\omega)$ is a smooth function. 
For the continuum band of the boson modes with $0<\tilde\omega_\nu<\omega_c$, we assume the spectral distribution having the form
\begin{align}
J(\omega)=
\begin{cases}
2\pi\alpha \omega^s/\omega_c^{s-1},\, {\rm for}\, 0<\omega< \omega_c;\\
0, {\rm otherwise}.
\end{cases}\label{j}
\end{align}
The overall coupling of the particles to the boson bath is characterized by the dimensionless parameter $\alpha$; this form of $J(\omega)$ has been widely studied in the spin-boson model \cite{leggett1987dynamics,weiss2012quantum}, in the context of which $s=1$ is called Ohmic with $s>1$ and $s<1$ called super-Ohmic and sub-Ohmic respectively. 
Analogous to the study of the spin-boson model \cite{leggett1987dynamics}, 
we are interested in how the particle dynamics at \emph{low} energy scale $\sim\Delta$ is influenced by the broad boson band whose \emph{high} energy cutoff  is $\omega_c$; we focus on the regime $\Delta/\omega_c\ll1$. To be specific, our following calculation takes $\omega_c$ as the energy unit, i.e., setting $\omega_c=1$, and assumes $\Delta=0.05$.

The key dynamics observable is the probability that the particles occupy the excited states, i.e., $P_e(t)\equiv\langle\hat N_e(t)\rangle/N$ with $\hat N_e\equiv \sum_{j=1}^N (\sigma^z_j+1)/2$. The dynamics starts from the initial state that all the particles are in the excited state and the boson bath is not excited, i.e., $\ket {\psi_N(0)}=\prod_{j=1}^N\ket e_j \otimes \ket 0$, where $\ket 0$ is the vacuum state of the boson bath. In the following,
we employ the numerical renormalization group method to calculate $P_e(t)$ \cite{anders2006spin}, 
and demonstrate that there exists a critical value $\alpha_c$ such that for long time, $P_e(t)$ decays to zero for $\alpha<\alpha_c$, and converges to a nonzero value for $\alpha>\alpha_c$. We show how $\alpha_c$ changes with the particle number $N$ and the interaction strength $g$.

\section{Dynamical transition}
A dynamical transition in $P_e(t)$ for $N=1$ can be shown analytically (see Appendix A). In this case, 
when in the absence of $H_{p-b}$, i.e., $\alpha=0$, the energy level of the initial state $\ket {\psi_1(0)}=\ket e \otimes\ket 0$, i.e., $\Delta/2$, is within the energy band of the continuum states $\{\ket{g}\otimes a_{\nu}^\dagger\ket{0}\}$, which spans from $-\Delta/2$ to $-\Delta/2+\omega_c$. None zero $\alpha$ couples the former with the latter. For sufficiently small $\alpha$, the eigen-energies of the full Hamiltonian form a continuum band, and $P_e(t)$ decays monotonically towards zero when $t\to\infty$ as the continuum band interferes distructively. However, when $\alpha$ exceeds a critical value $\alpha_c$, the original state $\ket {\psi_1(0)}$ is dressed by the boson modes so much, giving rise to a new discrete state $|D\rangle$ as the ground state below the bottom of the continuum band, and $P_e(t)$ converges to a nonzero value equal to $|\langle e|D\rangle\langle D\ket {\psi_1(0)}|^2$. For the spectral distribution, Eq.~(\ref{j}), one finds $\alpha_c=s\Delta/2\omega_c$.
The emergence of the discrete state $|D\rangle$ at the continuum band bottom is the underlying mechanism of the transition between $P_e(t\to\infty)=0$ and $P_e(t\to\infty)>0$.

\begin{figure*}
	\centering
	\includegraphics[width=1\textwidth]{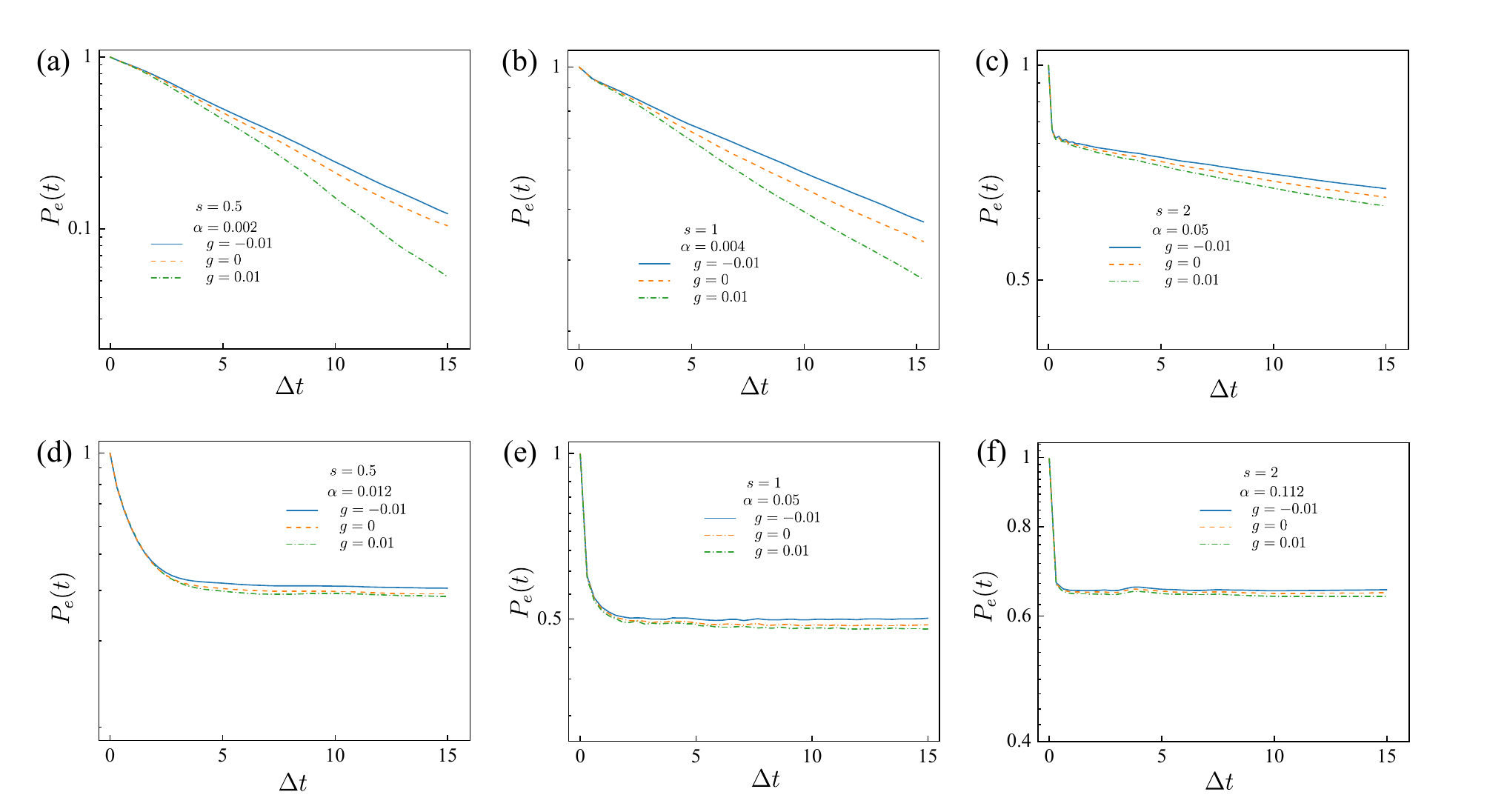}
	\caption{Numerical results of $P_e(t)$ for $N=2$ particles coupled to the boson bath, with $s=1/2,1,2$, $g=0,\pm0.01$ and various $\alpha$. 
	(a)-(c) for relatively small $\alpha$, $P_e(t)$ decays monotonically to zero. (d)-(f) when $\alpha$ is sufficiently large, $P_e(t)$ converges to a nonzero value for long time. The time evolution of $P_e(t)$ indicates the existence of a dynamical transition.}
	\label{2pe}
\end{figure*}

\begin{figure*}
	\centering
	\includegraphics[width=1\textwidth]{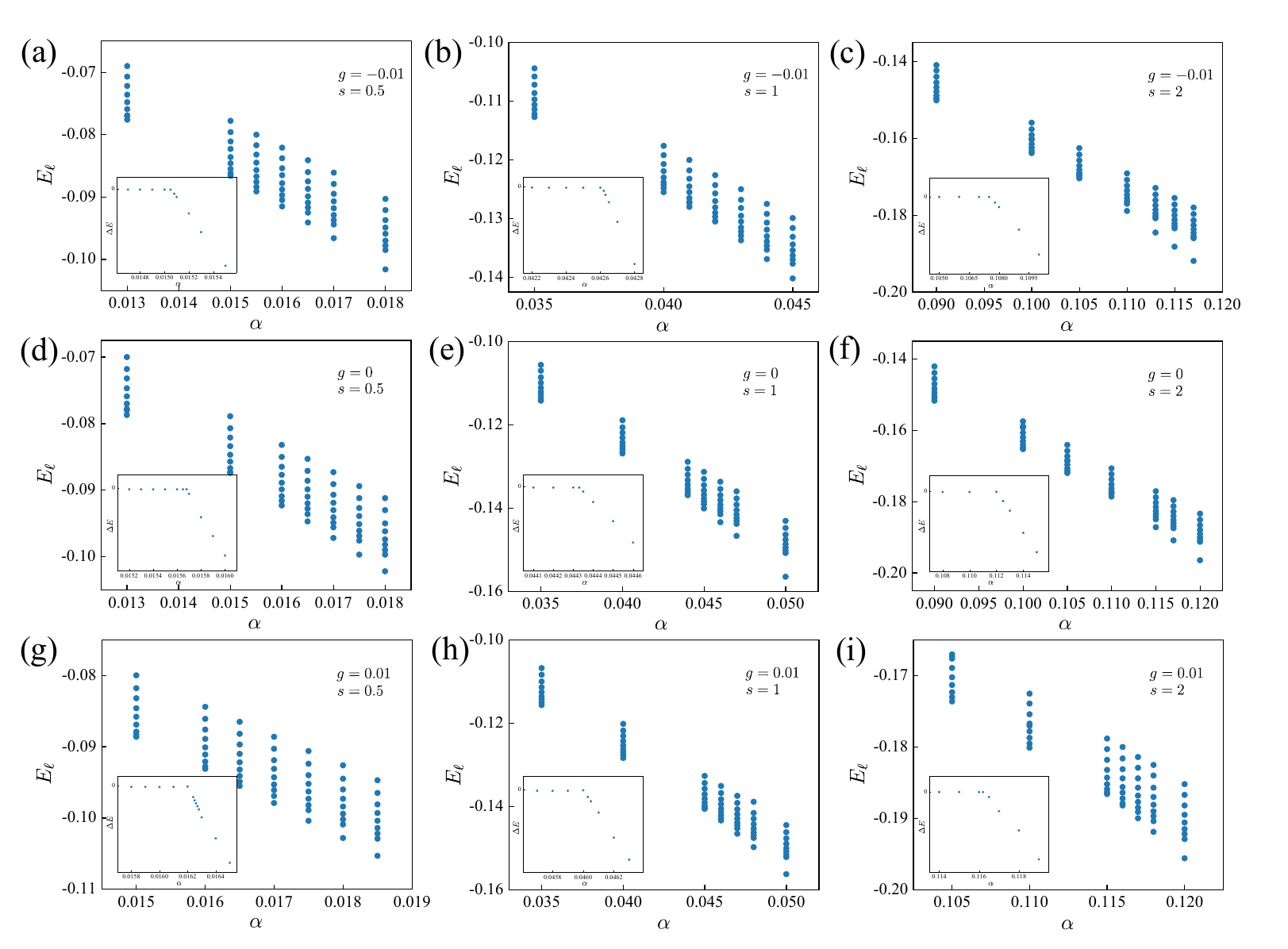}
	\caption{Numerical results of the lowest nine eigen-energies $E_\ell$ of $N=2$ particles coupled to the boson bath, with $s=1/2,1,2$, $g=0,\pm0.01$ and tuning $\alpha$. When $\alpha$ is relatively small, these eigen-energies cluster together, constituting the lowest part of a continuum band. The small gaps between the eigen-energies are due to the discretization of the continuum boson modes in the numerical algorithm. After $\alpha$ exceeds a critical value, the ground state eigen-energy $E_0$ separates from the bottom of the cluster, meaning that the ground state becomes a discrete state. The emergence of the discrete ground state gives rise to the dynamical transition in $P_e(t)$. The insets plot the energy difference $\Delta E\equiv E_0-E_1$ versus $\alpha$. By pinning down where $\Delta E$ turns nonzero, one can determine the critical value $\alpha_c$.} 
	 	\label{en}
\end{figure*}

For $N\ge2$, we employ the numerical renormalization group method to calculate the eigenstates $\{|\ell\rangle\}$ and eigenenergies $\{E_\ell\}$ of the Hamiltonian, Eq.~(\ref{h}), in the subspace that the conserved quantity $\mathcal C\equiv\sum_{j=1}^N |e_j\rangle\langle e_j|+\sum_{\nu} a_\nu^\dagger a_\nu$ equals $N$ (as the initial state $\ket {\psi_N(0)}=\prod_{j=1}^N\ket e_j \otimes \ket 0$ is an eigenstate of $\mathcal C$ with eigenvalue $N$, and $[\tilde H,\mathcal C]=0$). We label the ground state by $|\ell=0\rangle$. 
Our calculation follows the algorithm of the numerical renormalization group method prescribed for boson baths \cite{bulla2005numerical}. 
We take the discretization parameter $\Lambda=1.1$ to logarithmically discretize the continuum of the boson modes, and keep up to $N_S  = 1000$ states in each iteration, and usually run the iteration times up to $M=100$; the values of $\Lambda$, $N_S$ and $M$ give satisfactory convergence of numerical results (e.g., see Fig.~\ref{convergence}). 
An outline of our algorithm is provided in Appendix B. 
By the numerically calculated eigenstates $\{|\ell\rangle\}$ and eigenenergies $\{E_\ell\}$ of Eq.~(\ref{h}), we obtain
\begin{align}
P_e(t)=\sum_{\ell,\ell'}e^{-i(E_{\ell}-E_{\ell'})t}\langle\psi_N(0)|\ell'\rangle\langle\ell' |\hat N_e |\ell\rangle\langle \ell |\psi_N(0)\rangle/N\label{pet}.
\end{align}

Figure \ref{2pe} shows the numerical results of $P_e(t)$ for $N=2$, $\Delta=0.05$ and various 
$\alpha$, $s$ and $g$. 
In Fig.~\ref{2pe}(a)-(c), when $\alpha$ is relatively small, $P_e(t)$ decays monotonically toward \emph{zero}.
In contrast, when $\alpha$ is sufficiently large, Fig.~\ref{2pe}(d)-(f) indicates that $P_e(t)$ converges instead to a \emph{nonzero} value for long time. Comparing the behaviors of $P_e(t)$ shown for small and large $\alpha$, there shall be a dynamical transition in between. 
We numerically calculated $P_e(t)$ up to $N=8$ and found similar behaviors.

As in the case of $N=1$, we attribute such a dynamical transition to the emergence of a discrete state as the ground state of the whole system in the subspace of $\mathcal C=N$. To confirm the attribution and to locate the critical value $\alpha_c$ for the transition of $N$ particles with inter-particle interaction strength $g$, 
we investigate the eigenstates $\{|\ell\rangle\}$ and eigen-energies $\{E_\ell\}$ of Eq.~(\ref{h}) with $\mathcal C=N$, which are calculated numerically. 

Figure \ref{en} shows the numerical results of the lowest nine eigen-energies for $N=2$. When $\alpha$ is relatively small, 
all these eigen-energies cluster together, reflecting that they constitute the lowest part of a continuum band. The small gaps between the eigen-energies are due to the discretization of the continuum boson modes in the numerical algorithm, and can be shown to decrease with the iteration times $M$ (cf.~Fig.~\ref{el1m}). When $\alpha$ exceeds a critical value, the ground state eigen-energy $E_0$ separates from the bottom of the cluster, 
which indicates that the ground state $|\ell=0\rangle$ has become a discrete state. The discrete nature of $|\ell=0\rangle$ can be further confirmed by showing that $|\langle\ell=0|\psi_2(0)\rangle|^2$ converges to a nonzero value, while for the states $\ell\ge1$ remaining in the continuum, $|\langle\ell |\psi_2(0)\rangle|^2$ continues decreasing toward zero with the iteration times $M$ (cf.~Fig.~\ref{overlap}). Once the ground state $|\ell=0\rangle$ emerges as a discrete state from the continuum, for $t\to\infty$, the ground state contributes a finite value to $P_e(t)$ whereas all the other continuum states interfere destructively with each other completely [see Eq.~(\ref{pet})]. 


The calculated eigen-energies provide a way to determine the critical value $\alpha_c$. The insets of Fig.~\ref{en} show the energy difference $\Delta E\equiv E_0-E_1$ for $N=2$ when $\alpha$ is varied. When $\alpha$ is small, $|\ell=0\rangle$ and $|\ell=1\rangle$ are the lowest two neighboring states at the bottom of the continuum band and $\Delta E\to0$ with the iteration times $M$. On the other hand, when $\alpha>\alpha_c$, $|\ell=1\rangle$ becomes the bottom of the band and the energy gap $|\Delta E|$ is finite; at the critical point $\alpha_c$, the gap $|\Delta E|$ reduces to zero. We linear fit $\Delta E$ for $\alpha$ larger than and close to $\alpha_c$, and pin down $\alpha_c$ at the point where the linear fit crosses zero (cf.~Fig.~\ref{fit}). 
It may be tempting to identify $\alpha_c$ directly from the plots of $P_e(t)$ (e.g., Fig.~\ref{2pe}). However, the transition is defined by the value of $P_e(t)$ at $t\to\infty$. The reliability of inferring $P_e(t\to\infty)$ from the behavior of $P_e(t)$ within a finite duration can be questionable (see Appendix C for more discussions).

Table \ref{2ac} lists the value of $\alpha_c$ for $N=2$ determined from the insets of Fig.~\ref{en}. 
We find that for fixed $g$, $\alpha_c$ increases with $s$, as in the $N=1$ case where $\alpha_c=s\Delta/2\omega_c$. 
We also find that for fixed $s$, $\alpha_c$ increases with $g$; a ferromagnetic Ising coupling ($g<0$) reduces $\alpha_c$ and facilitates the transition. 

The dependence of $\alpha_c$ on $g$ can be understood by extrapolation from the small $\alpha$ limit, where one may picture the time evolution as the initial state decays in a cascade way, i.e., $|\psi_2(0)\rangle=|ee\rangle\otimes|0\rangle\to \{ |\psi_\nu\rangle\equiv\frac1{\sqrt2} (|ge\rangle+|eg\rangle) \otimes a_\nu^\dagger|0\rangle\}\to \{|\phi_{\nu\mu}\rangle\equiv|gg\rangle\otimes a_{\mu}^\dagger a_{\nu}^\dagger|0\rangle/\sqrt{1+\delta_{\nu\mu}}\}$. Considering that in the absence of $H_{p-b}$, i.e., $\alpha=0$, the energy difference between the unperturbed state $|ee\rangle\otimes|0\rangle$ and the band bottom of $|\psi_\nu\rangle$ is $\Delta+2g$, 
the Wigner-Weisskopf theory, applicable for small $\alpha$, would predict $P_e(t)\sim e^{-2J(\Delta+2g)t}$ during the first phase of the cascade [cf.~Eq.~(\ref{1wwa})]. Figure~\ref{2pe}(a)-(c) exhibit fairly good exponential decays, though discernible deviations from the exponential decays occur at small $t$. Similar deviations also appear in the case of $N=1$ (cf.~Fig.~\ref{oneatom}), and stem from the fact that the spectral function $J(\omega)$ is not constant, varying with $\omega$, and $J(\omega)$ at $\omega$ other than $\Delta+2g$ also contributes to $P_e(t)$ [cf.~Eq.~(\ref{uo})]. In the range $0<\omega<\omega_c$, as $s=0$ yields a constant $J(\omega)$, larger $s$ means $J(\omega)$ differing more from being constant; this is probably the reason why the early time deviation is the most obvious for $s=2$  when compared with $s=1/2$ and $1$. One may spot the zigzags at the beginning of the exponential decay phase in Fig.~\ref{2pe}(c), which is an artifact due to the discretization of the boson continuum and can be smoothed with finer discretization. 

Since $J(\omega)\sim\omega^s$, the decay rate $2J(\Delta+2g)$ shall increase with $g$, compatible with Fig.~\ref{2pe}(a)-(c). If the curves of $P_e(t)$ for different $g$, while all the other parameters are fixed (e.g., $s$), do not cross at any finite $t$ (we didn't find crossings in our numerical results), the behavior $P_e(t)\sim e^{-2J(\Delta+2g)t}$ predicted by the Wigner-Weisskopf theory indicates that $P_e(t,g_1)<P_e(t,g_2)$ for any $t$ if $g_1>g_2$. Now imagine that when we increase $\alpha$, the curves of $P_e(t,g)$ changes accordingly while the relation, $P_e(t,g_1)<P_e(t,g_2)$ if $g_1>g_2$, is maintained. Thus smaller $g$ shall 
have $P_e(t\to\infty)$ turning nonzero first and the transition takes place at smaller $\alpha$. Note that the above argument shall not be applicable to comparing $\alpha_c$ for different $s$; for $N=1$, we know that curves of $P_e(t)$ with different $s$ do cross each other [cf.~Eqs.~(\ref{1wwa}) and (\ref{pld})].

\begin{table}
\caption{Critical value $\alpha_c$ for $N=2$, $\Delta=0.05$ and various spectrum power $s$ and inter-particle interaction strength $g$.}\label{2ac}
\vspace{20pt}
\begin{tabular}{| c | c | c | c |}
\hline
$\alpha_c$ &  $s=1/2$ & $s=1$ & $s=2$ \\
\hline
$g=-0.01$ & $0.0151$ & $0.0426$ & $0.106$\\
\hline
$g=0$ & $0.0157$ & $0.0443$ & $0.112$ \\
\hline
$g=0.01$ & $0.0162$ & $0.046$ & $0.116$\\
\hline
\end{tabular}
\end{table}



\begin{figure}
	\centering
	\includegraphics[width=0.5 \textwidth]{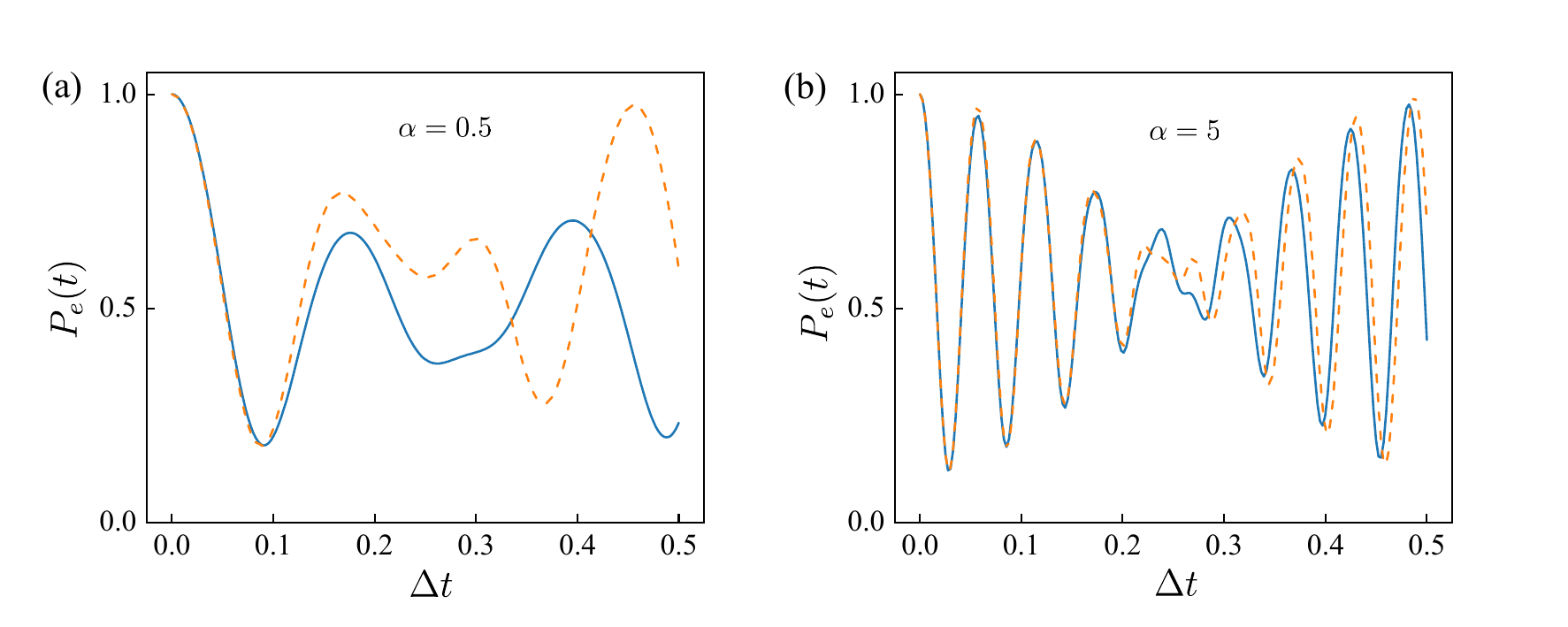}
	\caption{Numerical results of $P_e(t)$ for $N=2$, $s=1$, $g=0$ and $\alpha=0.5, 5$. For sufficiently large $\alpha$, $P_e(t)$ exhibits a notable oscillation. The dotted lines are produced from Eq.~(\ref{rabi}) combined with Eqs.~(\ref{epm}) and (\ref{e0}) for comparison; $P_e(t)$ of larger $\alpha$ agrees better with Eq.~(\ref{rabi}).}
	\label{biga}
\end{figure}

 
The existence of the dynamical transition is also supported by contrasting the $\alpha\to0$ and $\alpha\to\infty$ limits. 
Given Eq.~(\ref{j}), when $\alpha\to\infty$, $H_{p-b}$ dominates in Eq.~(\ref{h}) as $\lambda_\nu\sim\sqrt\alpha$. For $N=2$, $H_{p-b}$ couples the initial state $|\psi_2(0)\rangle=|ee\rangle\otimes|0\rangle$ to $\frac1{\sqrt2} (|ge\rangle+|eg\rangle) \otimes A^\dagger|0\rangle$, and subsequently to $\frac1{\sqrt{2}}|gg\rangle \otimes (A^\dagger)^2|0\rangle$ with $A\equiv \sum_\nu\lambda_\nu a_\nu/\sqrt{\sum_\mu\lambda_\mu^2}$ ($[A,A^\dagger]=1$). Using the above three states as the basis,
\begin{align}
H_{p-b}=\sqrt {\sum_{\mu}\lambda_\mu^2 } \begin {bmatrix}
  0 & \sqrt{2} & 0 \\
  \sqrt{2} & 0 & 2 \\
  0 & 2 &0 
\end{bmatrix},
\end{align}
which is diagonalized by the following three states

\begin{align}
|\chi_\pm\rangle=\begin{pmatrix}
1/\sqrt{6} \\
\pm1/\sqrt{2} \\
1/\sqrt{3}
\end{pmatrix}
\hspace{0.0cm},
|\chi_0\rangle =\begin{pmatrix}
-\sqrt{2/3} \\
0 \\
1/\sqrt{3}
\end{pmatrix}.
\end{align}
Thus the time-dependent state vector becomes
\begin{align}
|\psi_2(t)\rangle=\left( e^{-i E_+ t}|\chi_+\rangle-2 e^{-i E_0 t}|\chi_0\rangle+e^{-i E_- t}|\chi_-\rangle\right)/\sqrt6.
\end{align}
We evaluate $E_{\pm,0}$ to order $\sim\alpha^0$ and find 
\begin{align}
E_\pm&=\langle\chi_\pm|\tilde H|\chi_\pm\rangle\nonumber\\
&=\pm\sqrt {6\sum_{\mu}\lambda_\mu^2 }-\frac{\Delta}6+\frac{7\sum_\nu\tilde\omega_\nu\lambda_\nu^2}{6\sum_\mu\lambda_\mu^2}\nonumber\\
&=\pm\sqrt{\frac{12\alpha}{s+1}}\omega_c-\frac{\Delta}6+\frac{7(s+1)\omega_c}{6(s+2)},\label{epm}
\end{align}
and
\begin{align}
E_0&=\langle\chi_0|\tilde H|\chi_0\rangle=\frac{\Delta}3+g+\frac{2\sum_\nu\tilde\omega_\nu\lambda_\nu^2}{3\sum_\mu\lambda_\mu^2}\nonumber\\
&=\frac{\Delta}3+g+\frac{2(s+1)\omega_c}{3(s+2)}.\label{e0}
\end{align}
In the last lines of Eqs.~(\ref{epm}) and (\ref{e0}), we have substituted in the results $\sum_\mu\lambda_\mu^2=2\alpha\omega_c^2/(s+1)$ and $\sum_\mu\tilde\omega_\mu\lambda_\mu^2=2\alpha\omega_c^3/(s+2)$ obtained from Eqs.~(\ref{j0}) and (\ref{j}).
Resultantly
\begin{align}
P_e(t)=&\frac1{36}\left|e^{-i(E_+-E_0)t}+e^{-i(E_--E_0)t}+4\right|^2\nonumber\\
&+\frac1{24}\left|e^{-i(E_+-E_0)t}-e^{-i(E_--E_0)t}\right|^2,\label{rabi}
\end{align}
whose long time average is \emph{nonzero}. Additionally, the difference between $|E_+-E_0|$ and $|E_--E_0|$ is expected to lead to a beating in the $\alpha\to\infty$ limit. In contrast, for $\alpha\to0$, the long time average of $P_e(t)$ is $0^+$ since $P_e(t)$ decays to \emph{zero} with $t$. There must be a transition between the two limits. Figure \ref{biga} plots the evolution of $P_e(t)$ calculated numerically for $\alpha=0.5$ and $5$, which shows better agreement with Eq.~(\ref{rabi}) combined with Eqs.~(\ref{epm}) and (\ref{e0})
when $\alpha$ becomes larger.

\begin{figure*}
	\centering
	\includegraphics[width=1 \textwidth]{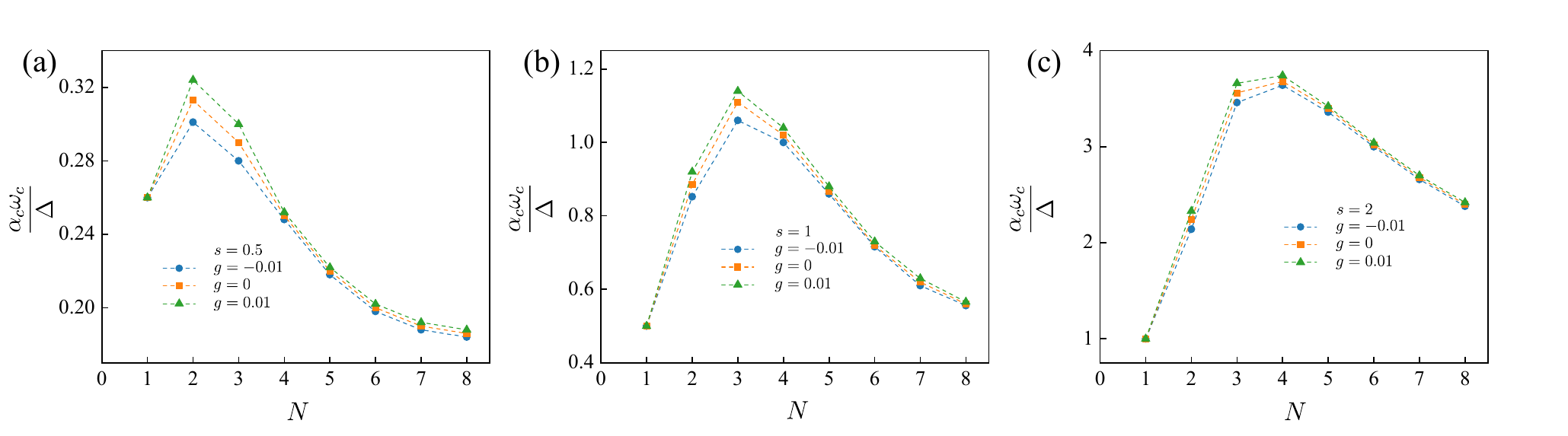}
	\caption{Dependence of the critical value $\alpha_c$ on $N$ and $g$ for $s=1/2,1,2$. The symbols are the numerical results. The lines are used to link the symbols. For $N=1$, the analytic result is $\alpha_c\omega_c/\Delta=s/2$. Our results show that $\alpha_c$ first increases and then decreases with $N$, and is suppressed (enlarged) by the $g<0$ ferromagnetic ($g>0$ anti-ferromagnetic) Ising interaction}
	\label{ac}
\end{figure*}

The numerical results of the cases that we calculated up to $N=8$ share similar qualitative features with those of $N=2$.
The above analysis detailed for $N=2$ can also be applied to general $N>1$. 
Our results of the critical value $\alpha_c$ for different values of $N$ and $g$ are summarized in 
Fig.~\ref{ac}. There is a maximum of $\alpha_c$ at finite $N$, which may be due to the competition between the following two factors. On the one hand, for simplicity, let us assume $g=0$; the energy difference between the unperturbed (also the initial) state $|\psi_N(0)\rangle=\prod_{j=1}^N|e_j\rangle\otimes|0\rangle$ and the unperturbed continuum bottom is $N\Delta$; larger $N$ makes it more difficult to dress the state $|\psi_N(0)\rangle$ into a stable discrete state below all the continuum states. On the other hand, if one discretizes the continuum of the boson modes into $\mathcal M$ discrete ones, the number of nonzero matrix elements of $H_{p-b}$ in the unperturbed basis would be $\sim \mathcal M+\mathcal M^2+\cdots+\mathcal M^N=\mathcal M(\mathcal M^N-1)/(\mathcal M-1)\approx e^{N\ln \mathcal M}$ (for $\mathcal M\gg1$), and these matrix elements are going to help to dress the state $|\psi_N(0)\rangle$ into the stable discrete state. As these matrix elements are relative smaller for larger $s$ since $J(\omega)$ decreases with $s$ for any $\omega\in[0,\omega_c]$, one expects that the maximum of $\alpha_c$ shall appear at larger $N$ for larger $s$, agreeing with Fig.~\ref{ac}.
The analysis of contrasting the $\alpha\to0$ and $\alpha\to\infty$ limits, as we apply above for $N=2$, indicates that for any finite $N$, there shall be a dynamical transition as $\alpha$ is varied. And Fig.~\ref{ac}(a) in particular suggests that $\alpha_c$ may converge to a finite value for $N\to\infty$.

\section{Discussion}
We studied the dynamics of the Jaynes–Cummings model generalized to multiple particles and a continuum boson bath. We also introduced an all to all Ising type inter-particle interaction. The dynamics starts with all the particles in their excited state and the boson bath in its vacuum state. The observable is the probability $P_e(t)$ that the particles remain in their excited state. We demonstrated that $P_e(t)$ exhibits a dynamical transition between decaying to zero and converging to a nonzero value when the system-environment coupling is tuned. We found that the critical coupling value exhibits a maximum at a finite $N$, and is suppressed (enlarged) by the ferromagnetic (anti-ferromagnetic) interaction. Our calculation is implemented via the non-perturbative numerical renormalization group method, and benchmarked with the case $N=1$ (Appendix C). Our results reveal how the number of particles and their intra-interaction affect the dynamical transition. 

The predictions of our study have the prospect to be tested experimentally by quantum simulation across multi-qubit platforms \cite{blatt2012quantum,graham2019rydberg,madjarov2020high,graham2022multi}. A larger number of qubits and a ferromagnetic Ising interaction are favorable to bring down the critical value $\alpha_c$ for the dynamical transition. Regarding the pairwise CNOT operation, if one encodes $|\uparrow\rangle=|e\rangle$ and $|\downarrow\rangle=|g\rangle$, our study implies that the state $|\uparrow\uparrow\rangle$ would suffer less dissipation if the CNOT operation is implemented by a ferromagnetic Ising interaction. We defer a detailed assessment of the CNOT gate performance in the generalized Jaynes–Cummings model to a future investigation.


\section*{Acknowledgements}
This work is supported by the National Natural Science Foundation of China Grant No.~12474270.

\section*{Appendix}








\subsection{Analytic results for $N=1$}\label{aa}
\begin{figure}
	\centering
	\includegraphics[width=0.5\textwidth]{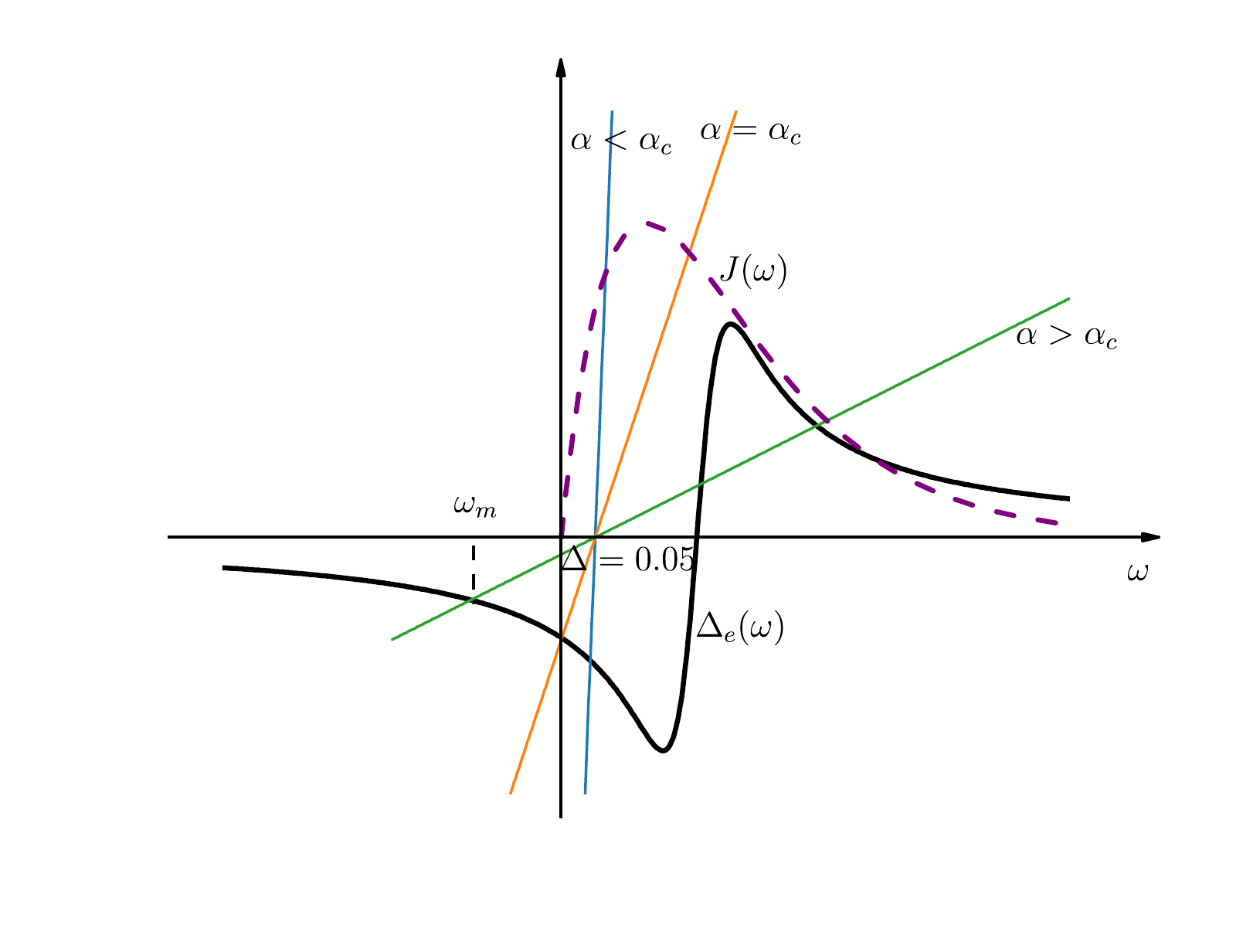}
	\caption{Generic spectral function $J(\omega)$, and its corresponding $\Delta_e(\omega)$ together with the straight line $(\omega-\Delta)/2\alpha$ for various $\alpha$. For sufficiently small $\alpha$, $\Delta_e(\omega)$ and the line $(\omega-\Delta)/2\alpha$ intersect at a frequency smaller than $\Delta$, which is denoted by $\omega_m$. As $\alpha$ increases, $\omega_m$ decreases. At the critical value $\alpha_c=-\Delta/2\Delta_e(0)$, $\omega_m=0$. For $\alpha>\alpha_c$, $\omega_m<0$, corresponding to a stable discrete dressed eigenstate with eigen-energy $\omega_m$ in the coupled system. It is this stable dressed state that gives rise to the nonzero value of $P_e(t)$ for long time in the case $N=1$.}
	\label{deltae}
\end{figure}

The existence of a dynamic transition for $N=1$ can be demonstrated analytically. 
In this case, $P_e(t)=|\mathcal{U}_e(t)|^2$ with $\mathcal{U}_e(t)\equiv\bra {\psi_1(0)} e^{-i\tilde Ht}\ket  {\psi_1(0)}$ and $\ket {\psi_1(0)}=\ket e \otimes\ket 0$. By 
the method of the Green's function \cite{cohen1998atom}, one can derive the Fourier transform $\mathcal{U}_e(\omega)$, defined by
\begin{align}
&\ \mathcal{U}_e(t)=\int_{-\infty}^\infty d \omega\ \mathcal{U}_e(\omega)\ e^{-i\omega t},&\label{ft}
\end{align}
having the form
\begin{equation}\begin{aligned}
&\mathcal{U}_e(\omega)=\lim_{\eta\to 0^{+}}\frac{1}\pi\ \frac{J(\omega)+\eta} {[\omega-\Delta-2\alpha \Delta_e(\omega)]^2+[J(\omega)+\eta]^2}.&\label{uo}
\end{aligned}\end{equation}
Here the spectral distribution $J(\omega)$ takes the role of the imaginary part of the self-energy, and correspondingly 
\begin{align}
\Delta_e(\omega)\equiv\mathcal P\int_{-\infty}^\infty \frac{d\omega'}{2\alpha\pi}\frac{J(\omega')}{\omega-\omega'}.\label{d}
\end{align}
Figure \ref{deltae} sketches a generic spectral distribution $J(\omega)$ [satisfying $J(\omega\to0^+)\to0^+$ and $J(\omega\to\infty)\to0^+$] and the resultant $\Delta_e(\omega)$.  

Combining Eqs.~(\ref{ft}) and (\ref{uo}), one can see that when $\alpha$ is sufficiently small, the contribution to Eq.~(\ref{ft}) is mainly from the frequency domain where $\omega-\Delta-2\alpha \Delta_e(\omega)\approx0$.
Figure \ref{deltae} shows the curve $\Delta_e(\omega)$ together with the straight line $(\omega-\Delta)/ 2\alpha$ of various $\alpha$ and correspondingly their intersections at frequency denoted by $\omega_m$, which is smaller than $\Delta$. Thus, for $\alpha\to0$, the domain $|\omega-\Delta|\lesssim J(\Delta)$ dominates the contribution to Eq.~(\ref{ft}), and one can approximate $\mathcal U_e(\omega)=
\frac{1}\pi\ \frac{J(\Delta)} {[\omega-\Delta-2\alpha \Delta_e(\Delta)]^2+J^2(\Delta)}$ as a Lorentzian, and obtain 
\begin{align}
P_e(t)=e^{-2 J(\Delta)t}, \label{1wwa}
\end{align}
as an exponential decay. However, it is known that for $t\to\infty$, the asymptote of $P_e(t)$ is a power law decay rather than Eq.~(\ref{1wwa}) \cite{cohen1998atom}. This is because when $t\to\infty$, the integral, Eq.~(\ref{ft}), is dominated by small $\omega$, i.e.,
\begin{align}
&\ \mathcal{U}_e(t)\sim\frac{2\alpha}{[\Delta+2\alpha\Delta_e(0)]^2 \omega_c^{s-1}}
\int_{0}^\infty d \omega \omega^s e^{-i\omega t},
\end{align}
where we have assumed that the low energy part of $J(\omega)$ can be approximated by $2\pi\alpha \omega^s/\omega_c^{s-1}$,
and resultantly
\begin{align}
P_e(t)\sim\left|\frac{2\alpha \Gamma(s+1)}{[\Delta+2\alpha\Delta_e(0)]^2 \omega_c^{s-1} t^{s+1}}\right|^2.\label{pld}
\end{align}
Nevertheless, $P_e(t)$ decays to zero as $t\to\infty$.

A qualitative change occurs at the critical point 
\begin{align}
\alpha_c=-\Delta/2\Delta_e(0), \label{alphac}
\end{align}
beyond which the straight line intersects $\Delta_e(\omega)$ at a negative frequency out of the range of the continuum. This intersection point corresponds to a discrete eigenstate with eigen-energy $\omega_m$ in the coupled system. 
Physically, the coupling has modified the initial discrete state of energy $\Delta$ so much that a dressed discrete state emerges below the continuum. This dressed discrete state is stable because it is not coupled to continuum states any more.

The emergence of the dressed discrete state brings about $\lim_{t\to\infty}P_e(t)\neq 0$. 
Since now for $\omega<0$, we have
\begin{equation}
\begin{aligned}
&\mathcal{U}_e(\omega)=\frac{1}{|1-2\alpha \Delta_m^{'}|}\delta(\omega-\omega_m),&\label{uem}
\end{aligned}\end{equation}
where $\Delta_m^{'}\equiv\partial_\omega \Delta_e |_{\omega=\omega_m}$. 
This delta function transforms to an undamped term, i.e., $\frac{1}{|1-2\alpha \Delta_m^{'}|}e^{-i\omega_m t}$, in $\mathcal U_e(t)$. For the spectral distribution, Eq.~(\ref{j}), we find $\Delta_e(0)=-\omega_c/s$ and $\alpha_c=s\Delta/2\omega_c$ 
consequently.

In the limit $\alpha\to\infty$, 
the leading term in $\tilde H$ becomes the coupling Hamiltonian
\begin{align}
&H_{p-b}=\sqrt{\sum_k \lambda_k^2} (\sigma^-A^\dagger+\sigma^+A),&
\end{align}
with $A\equiv\frac{1}{ \sqrt{\sum_k \lambda_k^2}}\sum_k \lambda_k a_k$ and $[A, A^\dagger]=1$.

In the basis \{$\ket e \otimes\ket 0$, $\ket g \otimes A^\dagger\ket 0$\}, this Hamiltonian can be expressed as
\begin{align}
	H_{p-b}=\sqrt {\sum_{\mu}\lambda_\mu^2 } \begin {bmatrix}
	0 & 1  \\
	1 & 0  
\end{bmatrix},
\end{align}
which is diagonalized by the following two states
\begin{align}
	|\pm\rangle=\begin{pmatrix}
		1/\sqrt{2} \\
		\pm1/\sqrt{2} 
	\end{pmatrix}.
\end{align}
Therefore, the time-dependent state vector is given by
\begin{align}
	|\psi_1(t)\rangle=\left( e^{-i E_+ t}|+\rangle+e^{-i E_- t}|-\rangle\right)/\sqrt2.
\end{align}
We calculate $E_{\pm,0}$ to order $\sim\alpha^0$ and obtain
\begin{align}
	E_\pm&=\langle\pm|\tilde H|\pm\rangle\nonumber\\
	&=\pm\sqrt {\sum_{\mu}\lambda_\mu^2 }+\frac{\sum_\nu\tilde\omega_\nu\lambda_\nu^2}{2\sum_\mu\lambda_\mu^2}\nonumber\\
	&=\pm\sqrt{\frac{2\alpha}{(s+1)}}\omega_c+\frac{(s+1)\omega_c}{2(s+2)}.\label{epm}
\end{align}
Starting from the initial state $\ket {\psi(0)}=\ket e \otimes\ket 0$ and assuming Eq.~(\ref{j}) for $J(\omega)$, we can work out 
\begin{align}
P_e(t)&=\frac{1}{4}\left|e^{-iE_+t}+e^{-iE_-t}\right|^2\nonumber\\
&=\cos^2\left(\sqrt{\frac{2\alpha}{s+1}}\omega_c t\right).\label{rabi1}
\end{align}
The qualitative difference in $P_e(t)$ between the small and large $\alpha$ also indicates that there must be a dynamical transition [see Eqs.~(\ref{1wwa}) and (\ref{rabi1})].

\subsection{Outline of the numerical renormalization group algorithm}\label{na}

\begin{figure}
	\centering
	\includegraphics[width=0.45\textwidth]{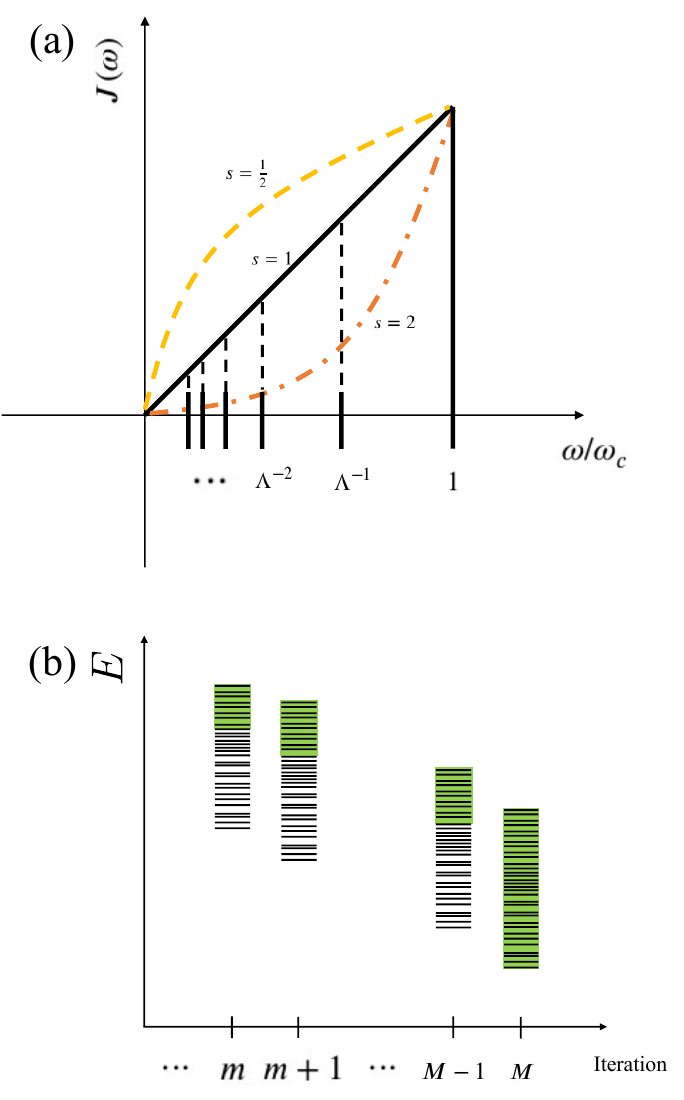}
	\caption{(a) Logarithmic discretization of the continuum boson modes with $\Lambda$ ($>1$) the dimensionless logarithmic discretization parameter. (b) Illustration of the iteration procedure of the numerical renormalization group method. The unshaded $N_S$ energy levels are fed into the next iteration while the shaded ones are left out. The iteration stops at $m=M$, and all the shaded states are taken as the eigen-states $|\ell\rangle$ of $\tilde H_M$ with $\ell=0,1,\dots$ labeling the states from low to high energy.}
	\label{fig. S1}
\end{figure}

The underlying mechanism of the dynamical transition is attributed to the emergence of a discrete state $|D\rangle$ at the continuum band bottom. The numerical renormalization group method is suitable to calculate the low energy levels of the system. Here we give an outline of the method. More details can be found in Ref.~\cite{bulla2005numerical}. 

To apply the method to the Hamiltonian, Eq.~(\ref{h}), with the continuum spectral function $J(\omega)$, Eq.~(\ref{j}), we discretize logarithmically the continuum of the bosonic bath modes as illustrated in Fig. \ref{fig. S1}(a), and $\Lambda$ is the dimensionless logarithmic discretization parameter chosen to be bigger than and close to $1$. After the discretization, the Hamiltonian retains its form, Eq.~(\ref{h}), whereas we use $a_\nu$ to denote the annihilation operator of the $\nu$th discretized bosonic mode ($\nu=0,1,\dots$), and correspondingly
\begin{align}
\lambda_\nu^2\approx& \int^{\omega_c/\Lambda^\nu}_{\omega_c/\Lambda^{\nu+1}} \frac{d\omega}{\pi } J(\omega)
= \frac{2\alpha\omega_c^2}{(s+1)\Lambda^{\nu(s+1)}} \left(1-\frac1{\Lambda^{s+1}}\right),\\
\tilde\omega_\nu\approx&\frac1{\lambda_\nu^2}\int^{\omega_c/\Lambda^\nu}_{\omega_c/\Lambda^{\nu+1}} \frac{d\omega}{\pi }J(\omega)\omega=\frac{(s+1)\omega_c}{(s+2)\Lambda^{\nu+1}}\frac{\Lambda^{s+2}-1}{\Lambda^{s+1}-1}.
\end{align}
Note that both $\lambda_\nu$ and $\omega_\nu$ decay exponentially with $\nu$; larger $\nu$ means closer to the continuum bottom. 

\begin{figure}
	\centering
	\includegraphics[width=0.5\textwidth]{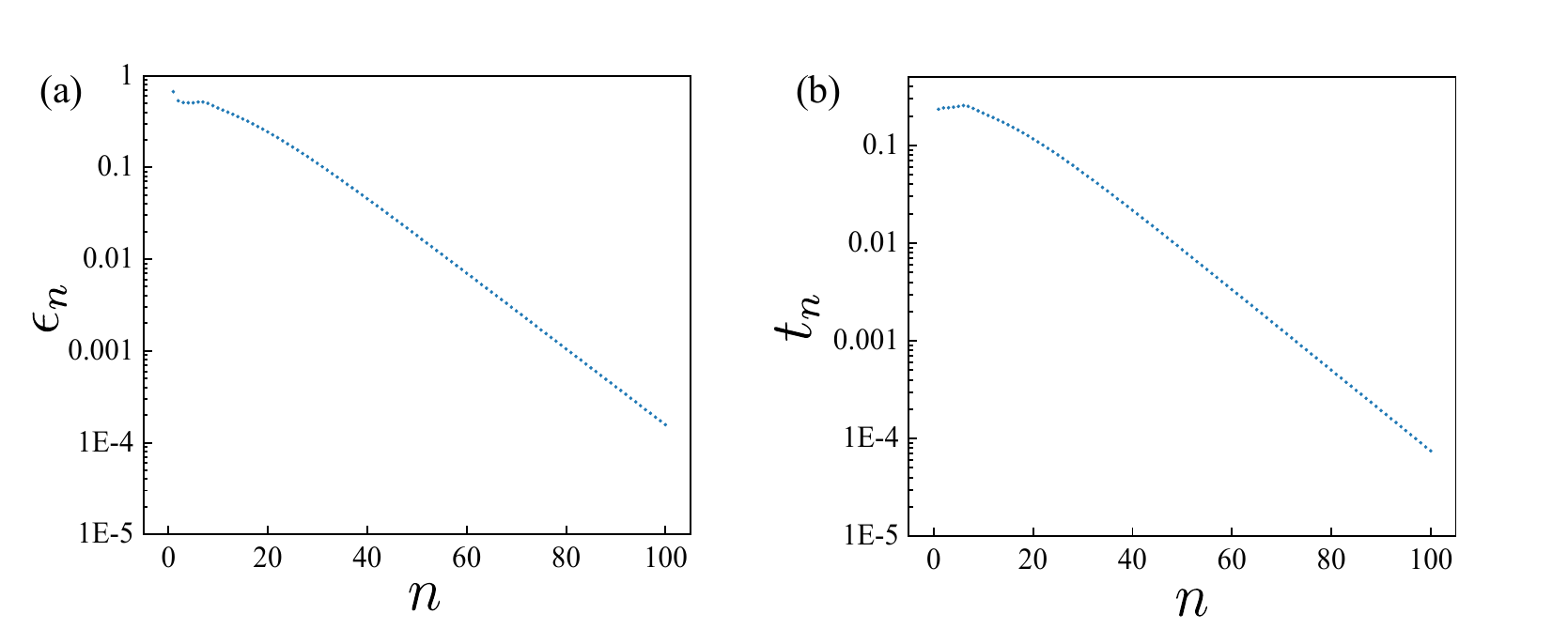}
	\caption{Plots of $\epsilon_n$ and $t_n$ for $\Lambda=1.1$, $\Delta=0.05$ and $\alpha=0.026$}
	\label{et}
\end{figure}

Further, there exists a real orthogonal transformation $U$ ($U^*=U$, $UU^T=U^TU=1$) which transforms $\{a_\nu, a_\nu^\dagger\}$ to a new set of bosonic modes $\{b_n, b_n^\dagger\}$ as
\begin{align}
b_n=&\sum_{\nu=0}^\infty U_{n\nu}a_{\nu},\\
a_{\nu}=&\sum_{n=0}^\infty U_{n\nu}b_n,
\end{align}
such that the Hamiltonian, Eq.~(\ref{h}), can be mapped into a Wilson chain:
\begin{align}
	\tilde H=&\sum_{j=1}^N\left[\frac{\Delta}2\sigma_j^z
	+\sqrt{\eta_0}\left(\sigma^-_jb_{0}^\dagger
	+\sigma^+_j b_0\right)\right]+g\sum_{j<k}\sigma_j^z \sigma_k^z\nonumber\\
	&+ \sum_{n=0}^{\infty} \left[ \epsilon_n b_n^\dagger b_n + t_n \left( b_n^\dagger b_{n+1} + b_{n+1}^\dagger b_n \right) \right]. \label{hb}
\end{align}
It is straightforward to show 
\begin{align}
\eta_0=&\sum_{\nu=0}^\infty \lambda_\nu^2=\int_0^{\omega_c}\frac{d\omega}{\pi} J(\omega),\\ 
U_{0\nu}=&\lambda_\nu/\sqrt{\eta_0},\\
\epsilon_0=&\sum_{\nu=0}^\infty\tilde \omega_\nu U_{0\nu}^2=\frac1{\eta_0}\int_0^{\omega_c}\frac{d\omega}{\pi}\omega J(\omega),\\
t_0=&\left[\frac1{\eta_0}\sum_{\nu=0}^\infty(\tilde\omega_\nu-\epsilon_0)^2\lambda_\nu^2\right]^{1/2},\\
U_{1\nu}=&(\tilde\omega_\nu-\epsilon_0)U_{0\nu}/t_0.
\end{align}
And the rest elements of $U$ can be determined iteratively by 
\begin{align}
U_{n+1,\nu}=[(\tilde\omega_\nu-\epsilon_n)U_{n\nu}-t_{n-1}U_{n-1,\nu}]/t_n,
\end{align}
together with
\begin{align}
\epsilon_n=&\sum_{\nu=0}^\infty \tilde\omega_\nu U_{n\nu}^2,\\
t_n=&\left\{\sum_{\nu=0}^\infty [(\tilde\omega_\nu-\epsilon_n)U_{n\nu}-t_{n-1}U_{n-1,\nu}]\right\}^{1/2}.
\end{align}
One important feature of $t_n$ and $\epsilon_n$ is that they also exponentially decay with $n$ (see Fig.~\ref{et}); the larger $n$, the finer energy scale is involved.

Thus, we use the energy levels of a truncated finite chain of length $m$, whose Hamiltonian is
\begin{align}
	\tilde H_m=&\sum_{j=1}^N\left[\frac{\Delta}2\sigma_j^z
	+\sqrt{\eta_0}\left(\sigma^-_jb_{0}^\dagger
	+\sigma^+_j b_0\right)\right]+g\sum_{j<k}\sigma_j^z \sigma_k^z\nonumber\\
	&+ \sum_{n=0}^{m}\epsilon_n b_n^\dagger b_n + \sum_{n=0}^{m-1}t_n \left( b_n^\dagger b_{n+1} + b_{n+1}^\dagger b_n \right),& \label{hm}
\end{align}
to approximate those of Eq.~(\ref{hb}). We can improve the approximation iteratively by adding an extra $b_{m+1}$ mode to the chain and re-diagonalize $\tilde H_{m+1}$ each time. However, to keep the dimension of the Hamiltonian to be re-diagonalized manageable, each time we retain at most the lowest $N_S$ levels for next iteration. We denote the eigenstates of $\tilde H_m$ as $|\ell;m\rangle$ with eigen-energy $E_{m,\ell}$; we use $\ell=0,1,2,\dots$ to label the energy levels from low to high.  
Suppose when we extend the finite chain to $m=m_0$, the dimension $D_{m_0}$ of $\tilde H_{m_0}$ exceeds $N_S$ for the first time. We assume the highest $D_{m_0}-N_S$ energy levels, i.e., $|\ell, m_0\rangle$ for $\ell=N_S,\dots, D_{m_0}-1$, accurate enough and leave them out of further iteration. On the other hand, we use the lowest $N_S$ energy levels and the Fock states of the $(m_0+1)$th bosonic mode $|n_{m_0+1}\rangle$ (satisfying $b^\dagger_{m_0+1}b_{m_0+1}|n_{m_0+1}\rangle=n_{m_0+1}|n_{m_0+1}\rangle$) to construct the basis
\begin{align}
|\ell, m_0\rangle\otimes |n_{m_0+1}\rangle,
\end{align}
with $\ell=0,1,\dots, N_S-1$ and $n_{m_0+1}=0,1,\dots, N_B$, and diagonalize $\tilde H_{m_0+1}$ in the constructed basis. 
From here, we repeat the above procedure all over again. Given the initial state $\ket {\psi_N(0)}=\prod_{j=1}^N\ket e_j \otimes \ket 0$, we are only interested in the subspace in which the conserved quantity $\mathcal C\equiv\sum_{j=1}^N |e_j\rangle\langle e_j|+\sum_{n=0}^\infty b_n^\dagger b_n$ equals $N$, and in accord only need to include bosonic mode Fock states $|n\rangle$ of occupation number $n$ up to $N$, i.e., choosing $N_B=N$. We carry out the iteration up to $m=M$ and the resultant energy levels $|\ell\rangle$ of energy $E_\ell$ are taken as the eigenstates of Eq.~(\ref{hb}). Figure \ref{fig. S1}(b) gives an illustration of the iteration procedure. 

\subsection{Benchmarking}

\begin{figure*}[htbp]
	\centering
	\includegraphics[width=1\textwidth]{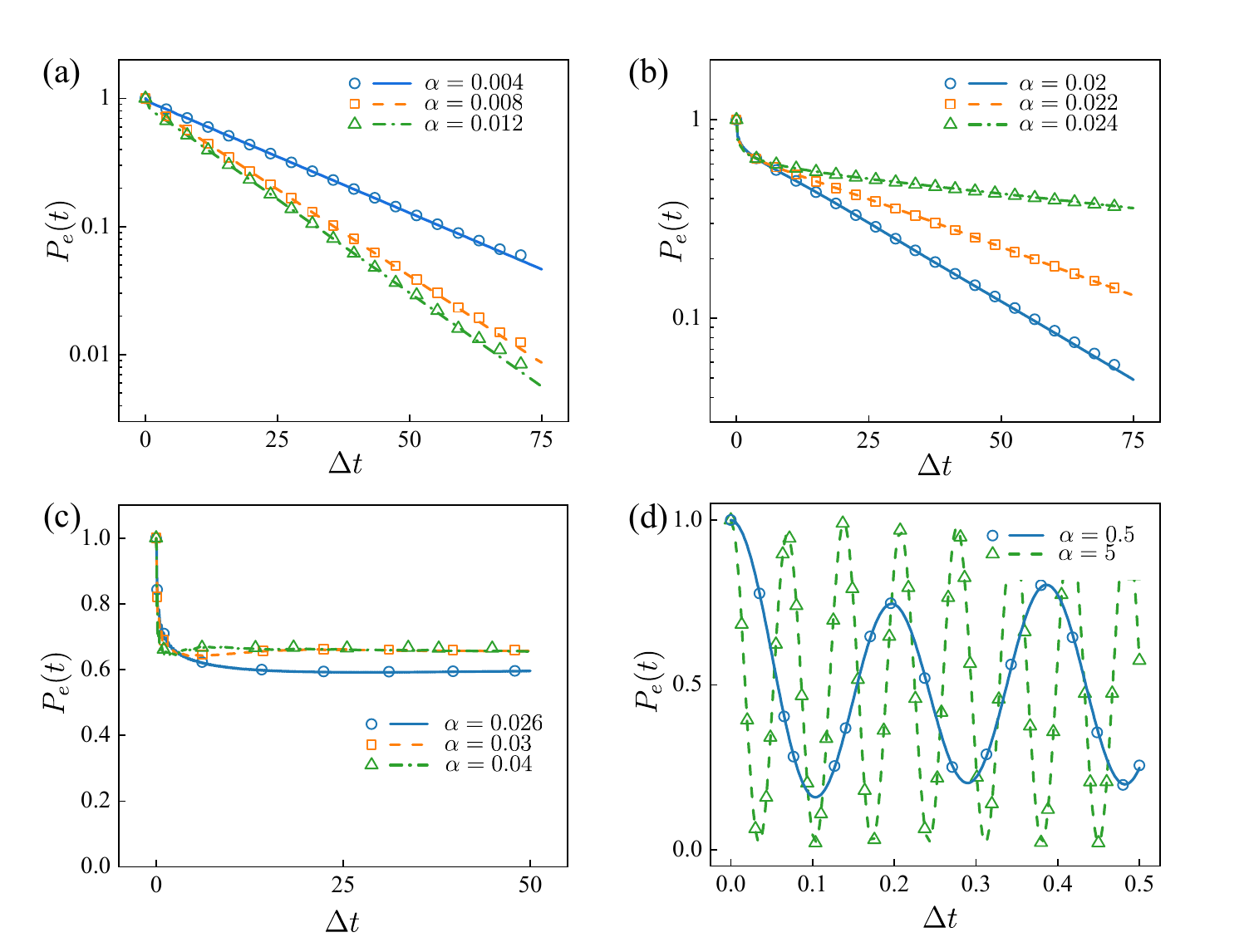}
	\caption{Time evolution of $P_e(t)$ for one single particle coupled to the boson bath. The symbols represent our numerical results. The lines are generated from Eq.~(\ref{uo}). For $\alpha$ below the critical value $\alpha_c=0.025$, (a) and (b) show that $P_e(t)$ decays to zero for long time. For $\alpha>\alpha_c$, (c) shows that $P_e(t)$ converges to a nonzero value for long time and an attenuating long time scale oscillation develops on top of the general trend of $P_e(t)$. This oscillation is enhanced for large $\alpha$ as shown in (d).}
	\label{oneatom}
\end{figure*}

The analytic results for $N=1$ provide a benchmark for our numerical calculation. In the case $N=1$, the initial state $\ket {\psi_1(0)}=\ket e \otimes\ket 0$ is coupled to the continuum of states $\{\ket{g}\otimes a_{\nu}^\dagger\ket{0}\}$. 
Figure \ref{oneatom} shows the results of $P_e(t)$ for $N=1$, $s=1$, $\Delta=0.05$ and various $\alpha$ by using Eq.~(\ref{j}); the discrete symbols are our numerical results, and the lines are generated from Eqs.~(\ref{ft}) to (\ref{d}).

For relatively small $\alpha$, Fig.~\ref{oneatom}(a) and (b) indicate that $P_e(t)$ decays towards zero monotonically over time. When $\alpha$ is sufficiently small, one assumes the perturbation theory applicable and expects an exponential decay of  $P_e(t)$ over time [see Eq.~(\ref{1wwa})], as shown in Fig.~\ref{oneatom}(a) and (b). As $\alpha$ increases, deviations from the exponential decays at small $t$ become more evident; this feature is due to the fact that $J(\omega)$ is not a constant \cite{cohen1998atom}.

Figure~\ref{oneatom}(c) and (d) show that when $\alpha$ is large enough, $P_e(t)$ behaves totally differently: $P_e(t)$ does not look to decay to zero any more for long time; phenomenologically it can be interpreted as a fraction of the particle's weight is trapped in the excited state. For even larger $\alpha$, Fig.~\ref{oneatom}(d) manifests an oscillation, whose origin is easily understood via the leading order of Eq.~(\ref{hb}) in the limit $\alpha\to\infty$ [see Eq.~(\ref{rabi1})]. 

The behavior of $P_e(t)$ shown in Fig.~\ref{oneatom} indicates a dynamic transition as $\alpha$ varies. 
The analytic result, Eq.~(\ref{alphac}), gives the critical value $\alpha_c=0.025$ for $s=1$ and $\Delta=0.05$.  
On the other hand, it is tempting to determine the value of $\alpha_c$ by inspecting directly the numerical results for $P_e(t)$. However, the difficulty lies in that the transition is defined by the value change of $P_e(t)$ at $t\to\infty$, while any numerical calculation is reliable up to certain finite time $t$. In fact, comparing Fig.~\ref{oneatom}(a)(b) with Eq.~(\ref{pld}), we conclude that our results of $P_e(t)$ have not well extend into the long time regime in which $P_e(t)$ shall decay in a power law yet. 
Therefore we turn to the energy levels calculated by the numerical renormalization group algorithm to determine $\alpha_c$.

\begin{figure}
	\centering
	\includegraphics[width=0.5\textwidth]{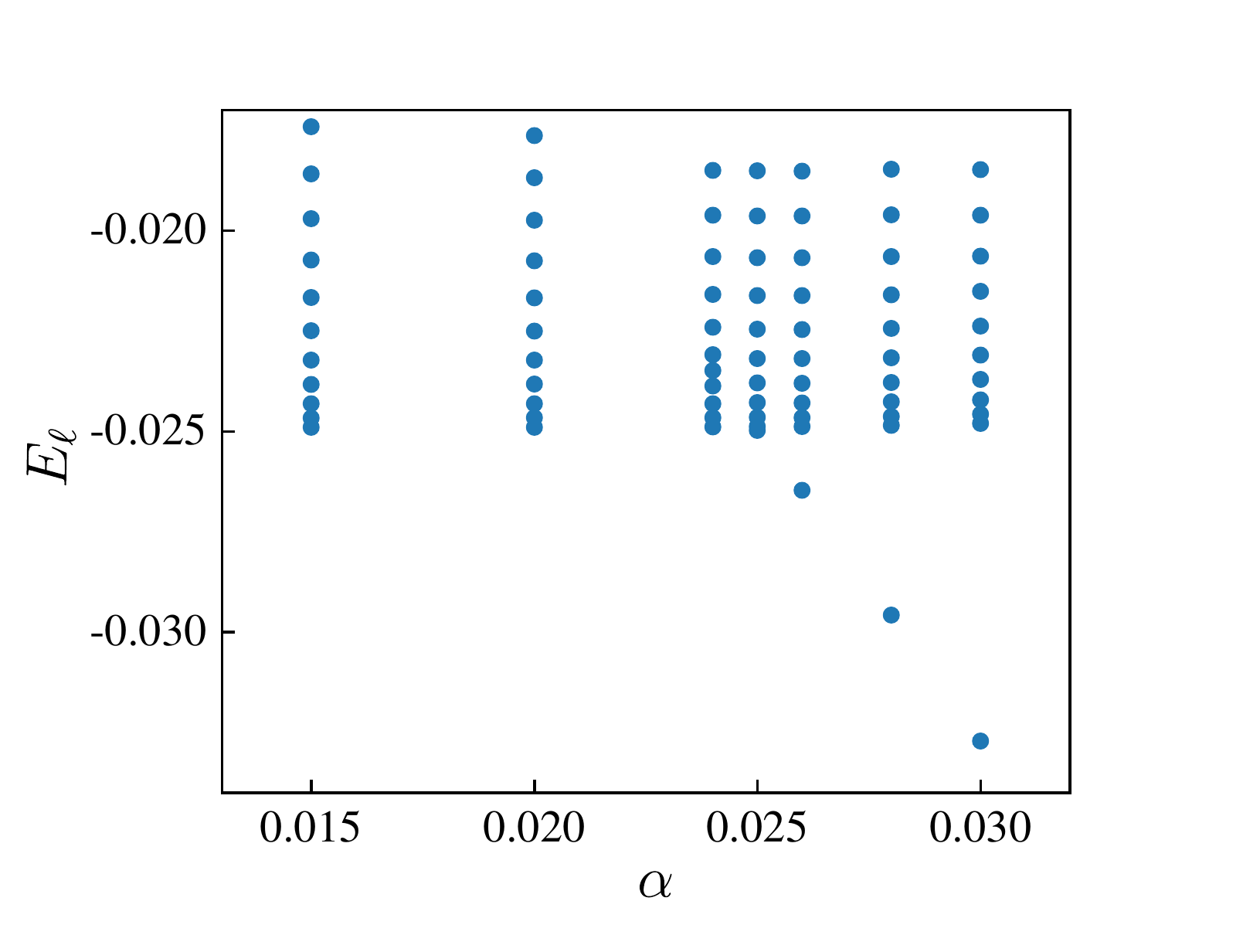}
	\caption{Numerical results of the lowest eleven eigen-energies $E_\ell$ of a single particle coupled to the boson bath, with $s=1$ and $\Delta=0.05$.}
	\label{el1alpha}
\end{figure}

\begin{figure}[htbp]
	\centering
	\includegraphics[width=0.5\textwidth]{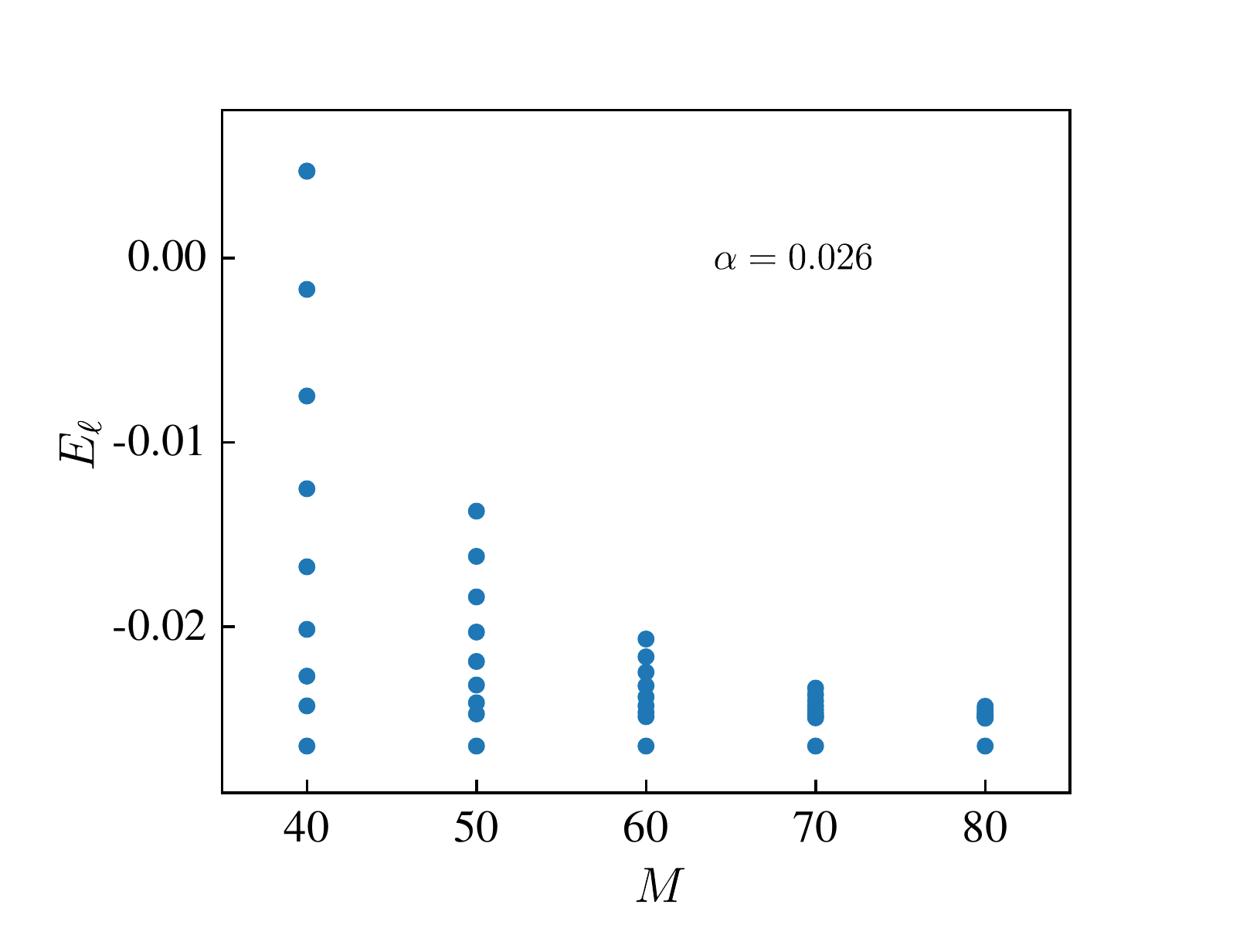}
	\caption{Variation of the lowest nine eigen-energies $E_\ell$ of a single particle coupled to the boson bath with the iteration times $M$ for $s=1$, $\Delta=0.05$ and $\alpha=0.026$ ($>\alpha_c=0.025$). With increasing $M$, the energy difference between states in a continuum band decreases towards zero, while the energy gap between a discrete state and the continuum band bottom converges to a finite value.}
	\label{el1m}
\end{figure}

\begin{figure}[htbp]
	\centering
	\includegraphics[width=0.5\textwidth]{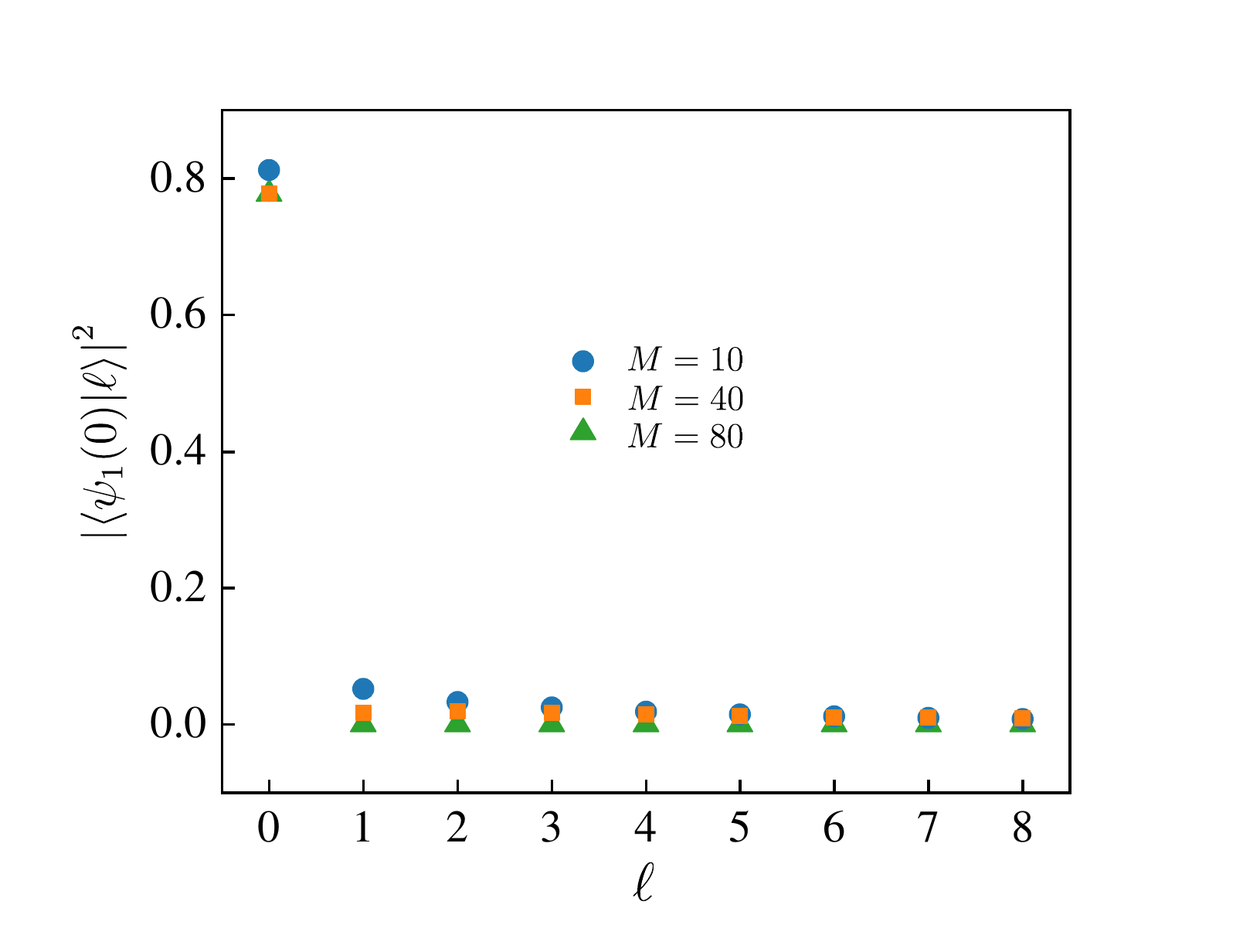}
	\caption{Variation of the overlap $|\langle\psi_1(0)|\ell\rangle|^2$ for the lowest nine eigen-states $|\ell\rangle$ of a single particle coupled to the boson bath with the iteration times $M$. The calculation is done with $s=1$, $\Delta=0.05$ and $\alpha=0.026$ ($>\alpha_c=0.025$). With increasing $M$, the overlap between the initial state $|\psi_1(0)\rangle$ and any continuum state $|\ell\rangle$ with $\ell\ge1$ decreases towards zero, while the overlap between the discrete ground state $|\ell=0\rangle$ and $|\psi_1(0)\rangle$ converges to a finite value.}
	\label{overlap}
\end{figure}

\begin{figure}[htbp]
	\centering
	\includegraphics[width=0.5\textwidth]{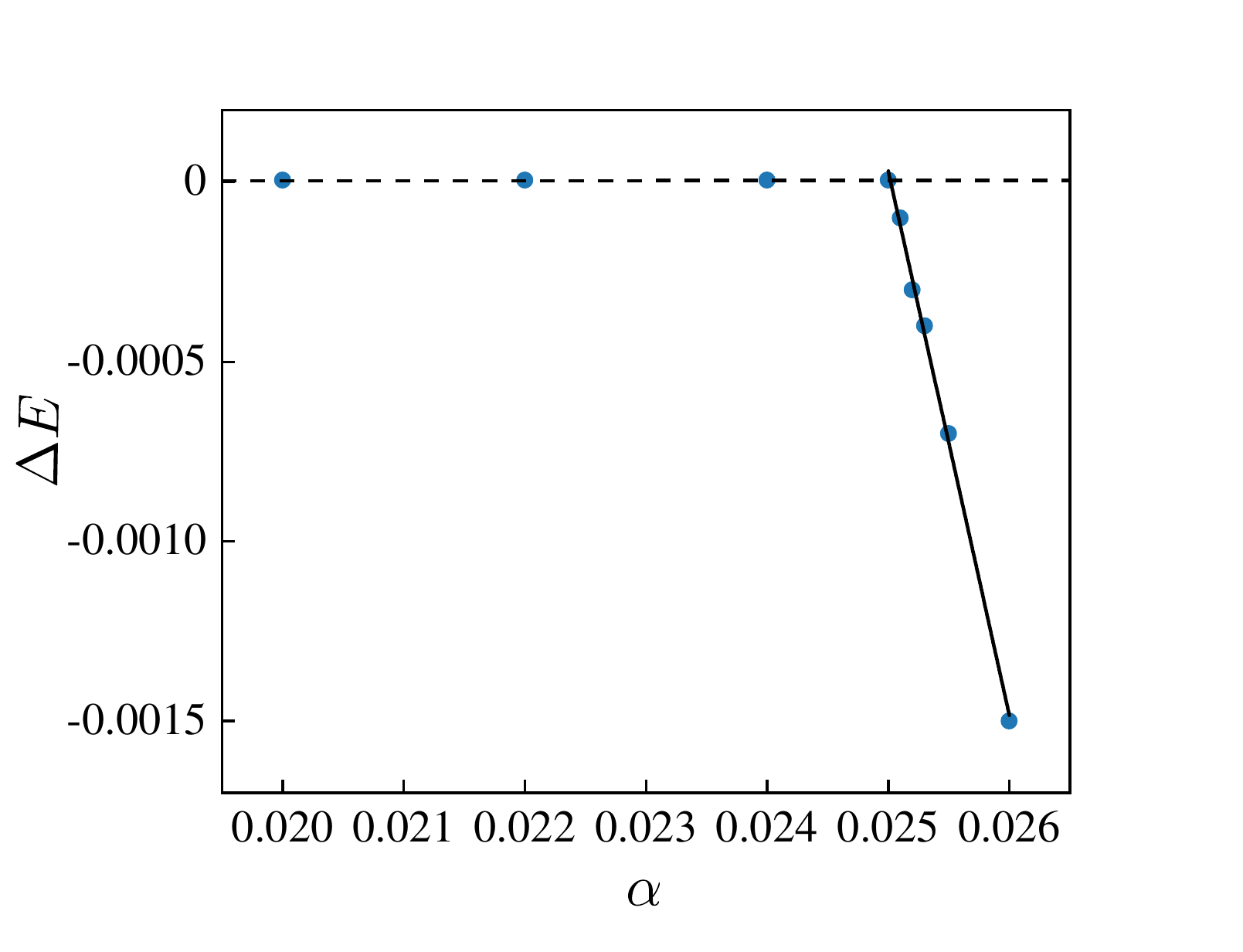}
	\caption{Numerical results of $\Delta E\equiv E_0-E_1$ of a single particle coupled to the boson bath with $s=1$ and $\Delta=0.05$. The solid line is a linear fit to the data points above $\alpha_c$. The dashed line represents $\Delta E=0$. The critical value $\alpha_c$ is nailed down at the crossing point between the two lines, agreeing well with the analytic result $\alpha_c=0.025$ for $\Delta=0.05$.}
	\label{fit}
\end{figure}

Figure \ref{el1alpha} plots the lowest eleven eigen-energies $E_\ell$ calculated numerically for $N=1$. These states form a cluster when $\alpha$ is small. The ground state $\ell=0$ separates from the cluster when $\alpha$ is sufficiently large. The states in the cluster are the continuum ones and once the ground state separates, it becomes a discrete state. This nature can be confirmed by inspecting how $E_\ell$ changes with the iteration times $M$. Figure \ref{el1m} shows that with increasing $M$, approaching finer energy resolution, the energy difference between the states in the cluster decreases, while $\Delta E\equiv E_0-E_1$ converges to a nonzero value. Furthermore, Fig.~\ref{overlap} shows that as expected, the probability $|\langle\psi_1(0)|\ell\rangle|^2$ diminishes for the continuum states and converges to a finite value for the discrete state as $M$ increases. Thus we determine the critical value $\alpha_c$ at the point where $\Delta E$ turns nonzero. Specifically speaking, as shown in Fig.~\ref{fit}, we linear fit $\Delta E$ for $\alpha$ larger than but close to $\alpha_c$, and pin point where the fit crosses zero. The crossing agrees well with the analytic result $\alpha_c=0.025$ for $\Delta=0.05$ and $s=1$.
 
\begin{figure}[htbp]
	\centering
	\includegraphics[width=0.5\textwidth]{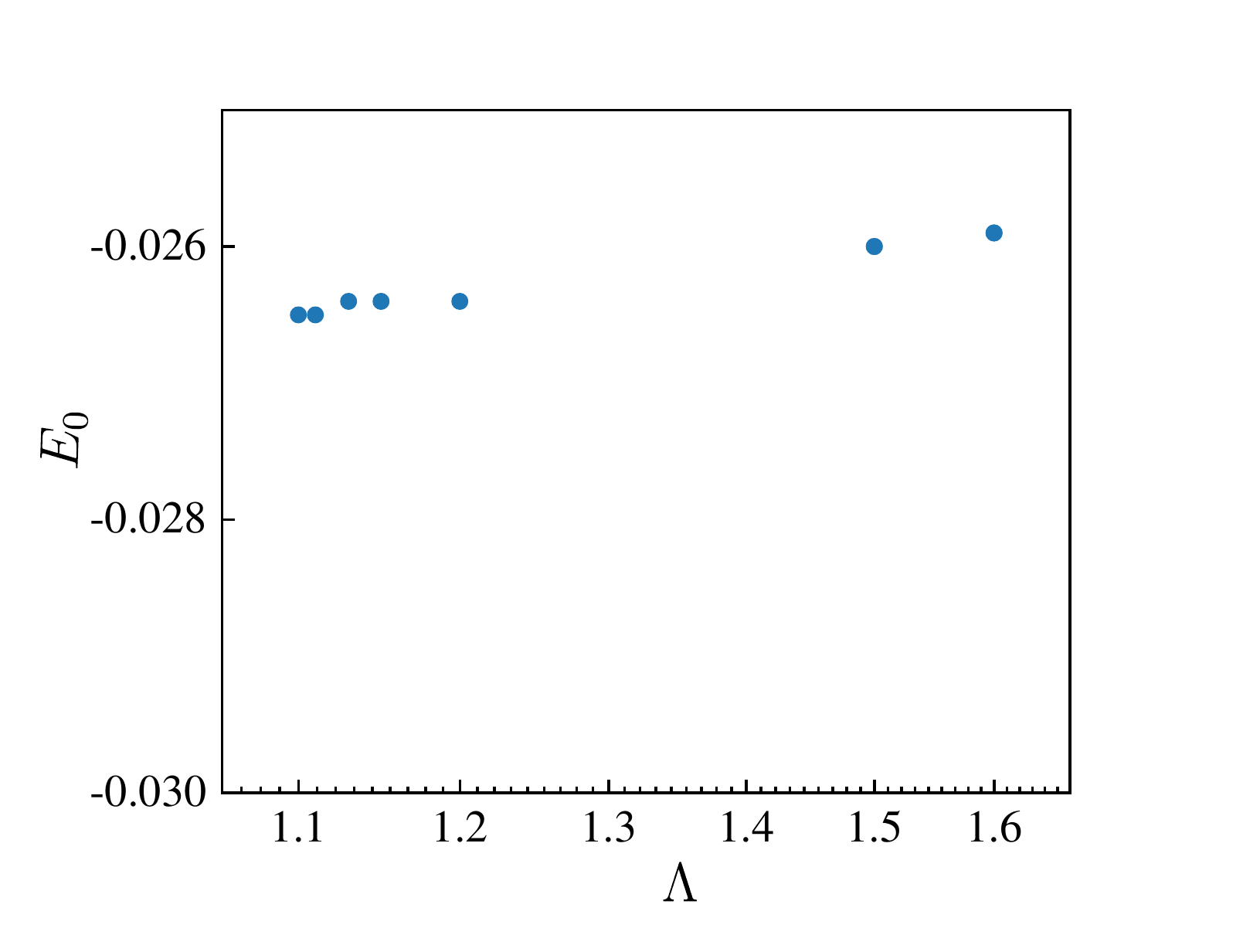}
	\includegraphics[width=0.5\textwidth]{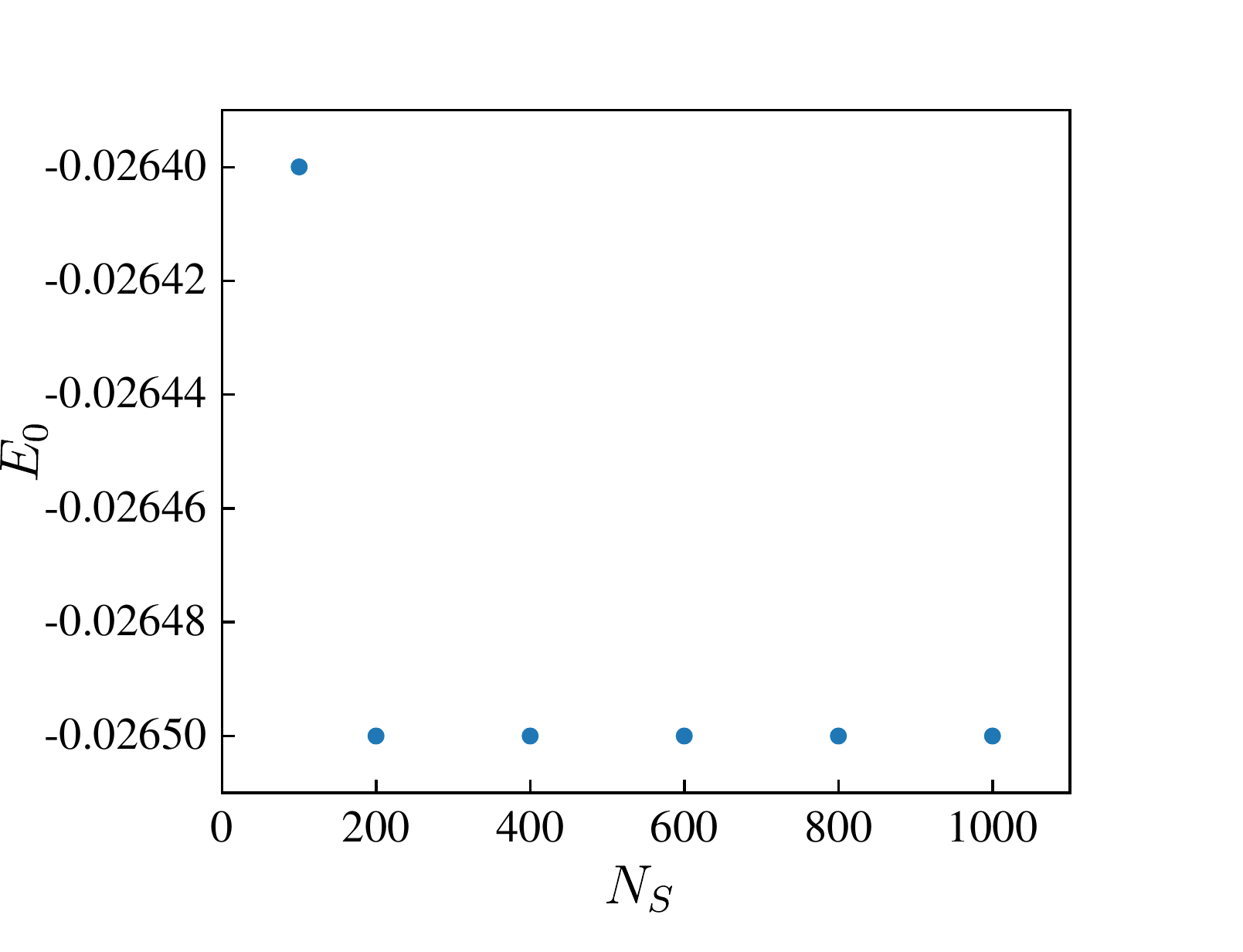}
	\includegraphics[width=0.5\textwidth]{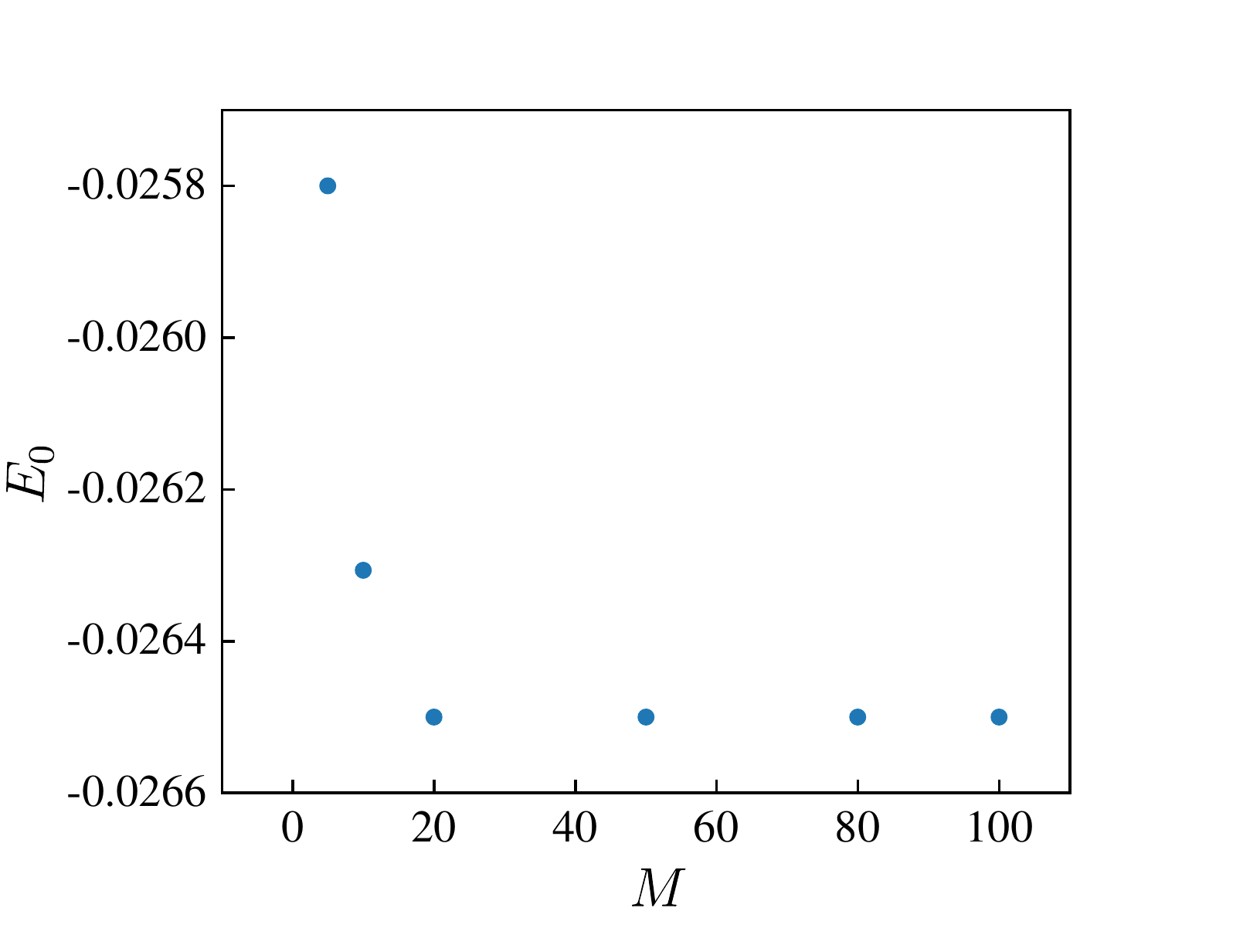}
	\caption{Convergence of $E_0$ with the logarithmic discretization parameter $\Lambda$ ($\to1$), the number of states kept for iteration $N_S$, and the iteration times $M$ for a single particle coupled to the boson bath with $s=1$, $\Delta=0.05$ and $\alpha=0.026$.}
	\label{convergence}
\end{figure}

Figure \ref{oneatom} shows that our numerical results of $P_e(t)$ agree well with the analytic calculation; the agreement benchmarks our numerical implementation of the numerical renormalization group algorithm. It is worth mentioning that all the parameters of the numerical algorithm are taken such that satisfactory convergence is met. Figure~\ref{convergence} demonstrates such convergence with the logarithmic discretization parameter $\Lambda$, the number of states kept for iteration $N_S$ and the iteration times $M$ via the physical quantity, the ground state energy $E_0$. Thus for our numerical results presented in the main text, we take the parameter values $\Lambda=1.1$, $N_S=1000$ and $M=100$ unless otherwise stated.  

\bibliographystyle{apsrev}
\bibliography{ref}
\end{document}